\documentclass[aps, prx, reprint, superscriptaddress, longbibliography, nourl, noeprint, nodoi]{revtex4-2}
\usepackage{amsmath,amssymb,physics,bm}
\usepackage{graphicx,microtype,siunitx} 
\usepackage{xcolor} 
\usepackage[colorlinks=true,linkcolor=blue,citecolor=blue,urlcolor=blue]{hyperref} 
\usepackage[all]{hypcap} 

\newcommand{\kB}[0]{k_{\rm B}}
\newcommand{\Dr}[0]{D_{\rm r}}

\begin{document} 
\title[]{Uphill and downhill first passage of an active Brownian particle: Asymmetry and exact path reweighting}

\author{Mykola Tasinkevych}
\email[]{mtasinkevych@ciencias.ulisboa.pt}
\affiliation{Departamento de F\'{\i}sica, Faculdade de Ci\^{e}ncias, Universidade de Lisboa, 1749-016 Lisboa, Portugal}
\affiliation{Centro de F\'{i}sica Te\'{o}rica e Computacional, Faculdade de Ci\^{e}ncias, Universidade de Lisboa, 1749-016 Lisboa, Portugal} 

\affiliation{International Institute for Sustainability with Knotted Chiral Meta Matter ($WPI$-$SKCM^2$), Hiroshima University, Higashi Hiroshima, Hiroshima 739-8531, Japan.}

\author{Xuan My Le}
\affiliation{Charles University, Faculty of Mathematics and Physics, Department of Macromolecular Physics, V Hole\v{s}ovi\v{c}k\'ach 2, CZ-18000 Praha 8, Czech Republic}

\author{Artem Ryabov}
\affiliation{Charles University, Faculty of Mathematics and Physics, Department of Macromolecular Physics, V Hole\v{s}ovi\v{c}k\'ach 2, CZ-18000 Praha 8, Czech Republic}

\date{\today}

\begin{abstract}
First passage processes in active systems combine
stochastic transport with self-propulsion and
orientational persistence, making motion along and
against an external bias sensitive to the internal active
dynamics. For passive biased diffusion, opposite exits can
have different splitting probabilities while their
conditional first passage time distributions remain
identical. We study how this relation changes for an
active Brownian particle driven by a constant external
force between two absorbing boundaries. Self-propulsion
breaks the equality of the uphill and downhill first
passage time distributions and modifies the splitting
probabilities. Nevertheless, the two directional path
ensembles remain exactly related by spatial reflection,
which pairs downhill and uphill first passage paths of the
same duration while preserving their orientational
history. The log-ratio of the probabilities of a path and
its reflected partner defines a path-dependent asymmetry
functional and provides an exact reweighting between the
two ensembles. In the symmetric half-weighted
representation, the uphill and downhill first passage time
distributions coincide for arbitrary orientational
persistence. The same path relation also allows rare
uphill statistics to be reconstructed from the more
frequently sampled downhill trajectories. Numerical simulations confirm the weighted equality across the
explored bias and persistence regimes, while perturbative and
asymptotic analyses clarify how orientational persistence produces
the directional asymmetry of the unweighted statistics. 
The exact relation between the two directional path ensembles suggests 
that similar symmetry-based reconstruction protocols may be found in other nonequilibrium first-passage problems.
\end{abstract}

\maketitle  

\section{Introduction}

Active Brownian particles (ABPs) provide a widely used
model of self-propelled motion. A particle moves with a
prescribed speed along its instantaneous
orientation, which changes through rotational diffusion,
while its position is also subject to translational
Brownian fluctuations
\cite{Marchetti2013,Bechinger2016}.
Self-propulsion provides a
sustained nonequilibrium drive that can substantially
modify transport under confinement or external forcing.
These effects are particularly relevant to escape,
transport, and target-search problems in confined or
biased environments, where the dynamics is often
characterised by the first arrival at a boundary or
target
\cite{CatesTailleur2015,Wang2016TargetSearch,
Woillez2019ActivatedEscape,Militaru2021EscapeDynamics}.
Recent studies have demonstrated the influence of active
dynamics on first passage and transition-path statistics
near absorbing boundaries and across barriers
\cite{Woillez2019ActivatedEscape,
Militaru2021EscapeDynamics,Caprini2019,
BaoucheEtAl2025,Caraglio2025}.
When an external bias selects a preferred direction, an
additional question arises: how are first passage events
along and against that direction related?

Direction-resolved first passage observables occur in many
stochastic transport problems with a preferred direction
generated, for example, by an external force, a
free-energy gradient, a chemical-potential difference, or
an imposed flow. In chemical and molecular physics,
transition-path and direction-conditioned timing
observables characterise barrier-crossing events and can
expose short-lived intermediates along folding and binding
pathways
\cite{Chung2012,Sturzenegger2018}.
In biophysics, they arise in particle and molecular
translocation through membrane channels
\cite{BerezhkovskiiHummerBezrukov2006}, forward and
backward stepping of molecular motors
\cite{Tsygankov2007}, and transport, target-search, and
encounter processes in confined or heterogeneous
environments
\cite{Condamin2007}.
In soft and condensed matter, direction-resolved first
passage observables describe motion along and against
applied fields or tilted energy landscapes
\cite{dagbug:2009}.
Two complementary pieces of information are then
available. Splitting probabilities quantify the relative likelihood
of the possible exit directions, whereas conditional
first passage time distributions characterise the durations of
trajectories associated with each exit. Differences
between these conditional distributions for opposite exit
directions can expose internal degrees of freedom, interactions, or nonequilibrium
driving that are less evident from splitting probabilities
or unconditional escape times alone
\cite{GladrowEtAl2019,BerezhkovskiiMakarov2019,
Tsygankov2007,preisler:2016}.

A remarkable result for biased Brownian diffusion is that the two exit directions can have different probabilities while their conditional first passage time distributions remain identical \cite{BerezhkovskiiHummerBezrukov2006,dagbug:2009}.
For diffusive particles traversing a membrane channel
between two reservoirs, the distributions of direct
translocation times were shown to be identical in the two
directions even in the presence of an asymmetric potential
across the channel
\cite{BerezhkovskiiHummerBezrukov2006}.
A closely related result was obtained for one-dimensional
diffusion in a periodic potential under a uniform driving
force: the time distributions for steps of one spatial
period along and against the force are identical, although
the probabilities of the two steps are different
\cite{dagbug:2009}.
Experiments with Brownian colloids in stationary optical
force landscapes subsequently confirmed corresponding
symmetries of exit-path and transition-path times and
showed that they can be broken by coloured nonequilibrium
forcing
\cite{GladrowEtAl2019}.
The sensitivity of this directional symmetry to
nonequilibrium forcing raises the question of how it is
modified by self-propulsion.

We address this question for an ABP driven by a constant
external force between two absorbing boundaries. The force
defines downhill and uphill first passage events according
to whether the particle exits along or against the bias.
Because the projection of the self-propulsion velocity
onto the force direction depends on the evolving
orientation, conditioning on opposite exits can select
trajectories with different orientational histories.

For the symmetric geometry considered here, the two
conditioned ensembles can be compared directly at the
level of individual paths. Under the spatial reflection
$X\to -X$, with the orientational history left unchanged,
every downhill first passage path is mapped onto an uphill
path of the same duration, and vice versa. The reflection
therefore defines a bijection between the two path
ensembles. Although the paths are paired geometrically,
they have different probabilities under the ABP dynamics.
Building on transition-path and path-weighting approaches
for stochastic systems
\cite{EVandenEijnden2010,Seifert2012}, we compare the
probabilities assigned to each path and its reflected
partner. Their log-ratio defines a path-dependent
asymmetry functional and leads to an exact reweighting
relation between the two directional ensembles.
Log-ratios of probabilities of conjugate paths provide
the basis for fluctuation relations and for reweighting
between different trajectory ensembles
\cite{Seifert2012,Crooks1999,Jarzynski1997,
DabelowBoEichhorn2019}. 
Relative path probabilities have also been extracted
directly from experimental trajectories
\cite{GladrowKeyserAdhikariKappler2021}, showing that
such path-probability information can be accessed
experimentally. In the present
problem, however, the conjugate paths are related by
spatial reflection rather than physical time reversal,
and no thermodynamic interpretation of the asymmetry
functional is required.

For the ABP, the asymmetry functional depends on the
orientational history, so the unweighted uphill and
downhill conditional first passage time distributions are
different. When the contribution of each path is
multiplied by the corresponding exponential factor
determined by the asymmetry functional, the reweighted
uphill and downhill first passage time distributions become
identical. The equality of the reweighted distributions
is exact and holds for any orientational persistence.
Thus, activity generates a directional asymmetry in the
observable first passage statistics, while spatial
reflection pairs downhill and uphill first passage paths
whose relative probabilities are determined exactly by
the path-dependent asymmetry functional.

One practical consequence of this equality is the
reconstruction of rare uphill statistics. In a biased
system, downhill crossings are sampled much more
frequently than uphill passages, so the better sampled
downhill trajectories can be used to recover the uphill
first passage time distribution by reweighting their
contributions with an exponential factor determined by
the asymmetry functional. At the same time, the
differences between the unweighted uphill and downhill
observables retain information about self-propulsion and
persistence. Together with the splitting probabilities,
these observables provide complementary measures of how
active dynamics modifies biased first passage transport.

The remainder of the paper is organised as follows. In
Sec.~\ref{sec:model} we introduce the ABP model, the
dimensionless control parameters, and the projected first
passage setting. In Sec.~\ref{sec:transition_paths} we
formulate the transition-path description and derive the
path-reweighting relation. Section~\ref{sec:survival_splitting}
presents the corresponding Fokker--Planck
description, the weak-persistence expansion, and the
frozen-orientation approximation, with the detailed
perturbative derivation and its numerical confirmation
given in Appendix~\ref{app:weak_persistence}.
In Sec.~\ref{sec:numerics} we test the path-reweighting relation
 numerically  and examine how the directional asymmetries
depend on self-propulsion and external bias. Finally,
Sec.~\ref{sec:experimental} discusses possible
experimental protocols and implications, and
Sec.~\ref{conclusion} concludes the paper.

\section{Model and projected first passage setting}
\label{sec:model}

\subsection{Planar active-particle dynamics and
first passage event}

We consider a single active Brownian particle moving in
two spatial dimensions under the action of a constant
external force $f$ directed along the $x$ axis. The
particle self-propels at a constant speed $u$ along its
instantaneous orientation
\[
\mathbf{e}(t)
=
\bigl(\cos\phi(t),\sin\phi(t)\bigr).
\]
Thermal fluctuations of the surrounding fluid at
temperature $T$ produce translational and rotational
diffusion. In the overdamped regime, the particle position
$(x,y)$ and orientation $\phi$ obey
\begin{subequations}
\label{eq:Langevin-SI}
\begin{align}
\frac{dx}{dt}
&=
u\cos\phi+\mu f+\sqrt{2D}\,\xi_x(t),
\label{eq:model_x}
\\
\frac{dy}{dt}
&=
u\sin\phi+\sqrt{2D}\,\xi_y(t),
\label{eq:model_y}
\\
\frac{d\phi}{dt}
&=
\sqrt{2\Dr}\,\xi_{\rm r}(t),
\label{eq:model_phi}
\end{align}
\end{subequations}
where $\mu$ is the mobility,
$D=\mu\kB T$ is the translational diffusion coefficient,
$\Dr$ is the rotational diffusion coefficient, and
$\kB$ is the Boltzmann constant. The noises
$\xi_x(t)$, $\xi_y(t)$, and $\xi_{\rm r}(t)$ are
independent Gaussian white-noise processes satisfying
\[
\langle\xi_\alpha(t)\rangle=0,
\qquad
\langle
\xi_\alpha(t)\xi_{\alpha'}(t')
\rangle
=
\delta_{\alpha\alpha'}\delta(t-t'),
\]
with $\alpha,\alpha'\in\{x,y,{\rm r}\}$.

Equations~\eqref{eq:model_x} and
\eqref{eq:model_phi} form a closed stochastic subsystem
for the position $x(t)$ along the force direction and the
orientation $\phi(t)$: neither variable depends on the
transverse coordinate $y(t)$. This property is important
for the first passage problem considered below, whose
stopping condition depends only on $x(t)$. Consequently,
all quantitative first passage observables are completely
determined by the reduced process $(x,\phi)$.
Equation~\eqref{eq:model_y} is integrated only to
generate the illustrative planar trajectories shown in
Fig.~\ref{fig_trajectories_2D}.

At the initial time $t=0$, the particle is placed at the
origin,
\begin{equation}
x(0)=y(0)=0,
\end{equation}
and the initial orientation $\phi(0)$ is drawn uniformly
from $[0,2\pi)$.

The dynamics is terminated when the particle first
reaches either of the absorbing boundaries
\begin{equation}
x=\pm\frac{L}{2}.
\end{equation}
For $f>0$, the external potential is
$V(x)=-fx$. A first arrival at $x=+L/2$ therefore
defines a downhill passage along the applied force,
whereas a first arrival at $x=-L/2$ defines an uphill
passage against it.

\subsection{Dimensionless formulation and control
parameters}

We introduce the dimensionless variables before
presenting the representative trajectories and
first passage observables. Using the boundary separation
$L$ as the unit of length and the rotational diffusion
time $\Dr^{-1}$ as the unit of time, we define
\begin{equation}
X=\frac{x}{L},
\qquad
Y=\frac{y}{L},
\qquad
\tau=\Dr t.
\end{equation}
The absorbing boundaries are then located at
$X=\pm1/2$, and Eqs.~\eqref{eq:Langevin-SI} become
\begin{subequations}
\label{eq:Langevin-dimless}
\begin{align}
\frac{dX}{d\tau}
&=
U\cos\phi+\widetilde D F
+\sqrt{2\widetilde D}\,\xi_x(\tau),
\label{eq:dimless_x}
\\
\frac{dY}{d\tau}
&=
U\sin\phi
+\sqrt{2\widetilde D}\,\xi_y(\tau),
\label{eq:dimless_y}
\\
\frac{d\phi}{d\tau}
&=
\sqrt{2}\,\xi_{\rm r}(\tau).
\label{eq:dimless_phi}
\end{align}
\end{subequations}
The rescaled noises remain independent Gaussian
white-noise processes with zero mean and unit variance.
The dimensionless control parameters are
\begin{equation}
U=\frac{u}{\Dr L},
\qquad
F=\frac{fL}{\kB T},
\qquad
\widetilde D=\frac{D}{\Dr L^2}.
\end{equation}

The dimensionless self-propulsion speed may be written as
\[
U=\frac{\ell_p}{L},
\qquad
\ell_p=\frac{u}{\Dr},
\]
where $\ell_p$ is the persistence length. Thus,
$U\ll1$ corresponds to weakly persistent motion on the
scale of the first passage interval, whereas
$U\gtrsim1$ corresponds to persistence lengths comparable
to or larger than the boundary separation.

The dimensionless force $F$ is the work performed by
the external force over the distance $L$, measured in
units of thermal energy. The dimensionless diffusivity
$\widetilde D$ measures the translational diffusion
accumulated during one orientational-relaxation time.
Introducing the corresponding diffusive length
\[
\ell_D=\sqrt{\frac{D}{\Dr}},
\]
one has
\[
\widetilde D
=
\left(\frac{\ell_D}{L}\right)^2.
\]
The dimensionless force-driven drift is the combination
\[
\widetilde D F
=
\frac{\mu f}{\Dr L}.
\]

This parametrisation is convenient for comparison with
experiments on active Janus colloids and related systems
\cite{Bechinger2016}. For a spherical particle of radius
$R$ in a fluid of viscosity $\eta$, the translational and
rotational diffusion coefficients are approximately
\begin{equation}
D=\frac{\kB T}{6\pi\eta R},
\qquad
\Dr=\frac{\kB T}{8\pi\eta R^3},
\end{equation}
which gives
\begin{equation}
\widetilde D=\frac{4R^2}{3L^2}.
\end{equation}
Once the particle radius and the experimental crossing
length are fixed, $\widetilde D$ is therefore determined
geometrically, while $U$ and $F$ can be varied through
the propulsion speed $u$ and the applied force $f$.

In the following analysis, we use
$\{U,F,\widetilde D\}$ as the natural set of control
parameters. All quantitative first passage calculations
are obtained from the closed
process $(X,\phi)$ governed by
Eqs.~\eqref{eq:dimless_x} and
\eqref{eq:dimless_phi}. The transverse equation
\eqref{eq:dimless_y} is used only for the planar
visualisation in Fig.~\ref{fig_trajectories_2D}.

\subsection{Representative trajectories and
first passage observables}

\begin{figure*}[t]
    \centering
    \includegraphics[width=0.8\textwidth]{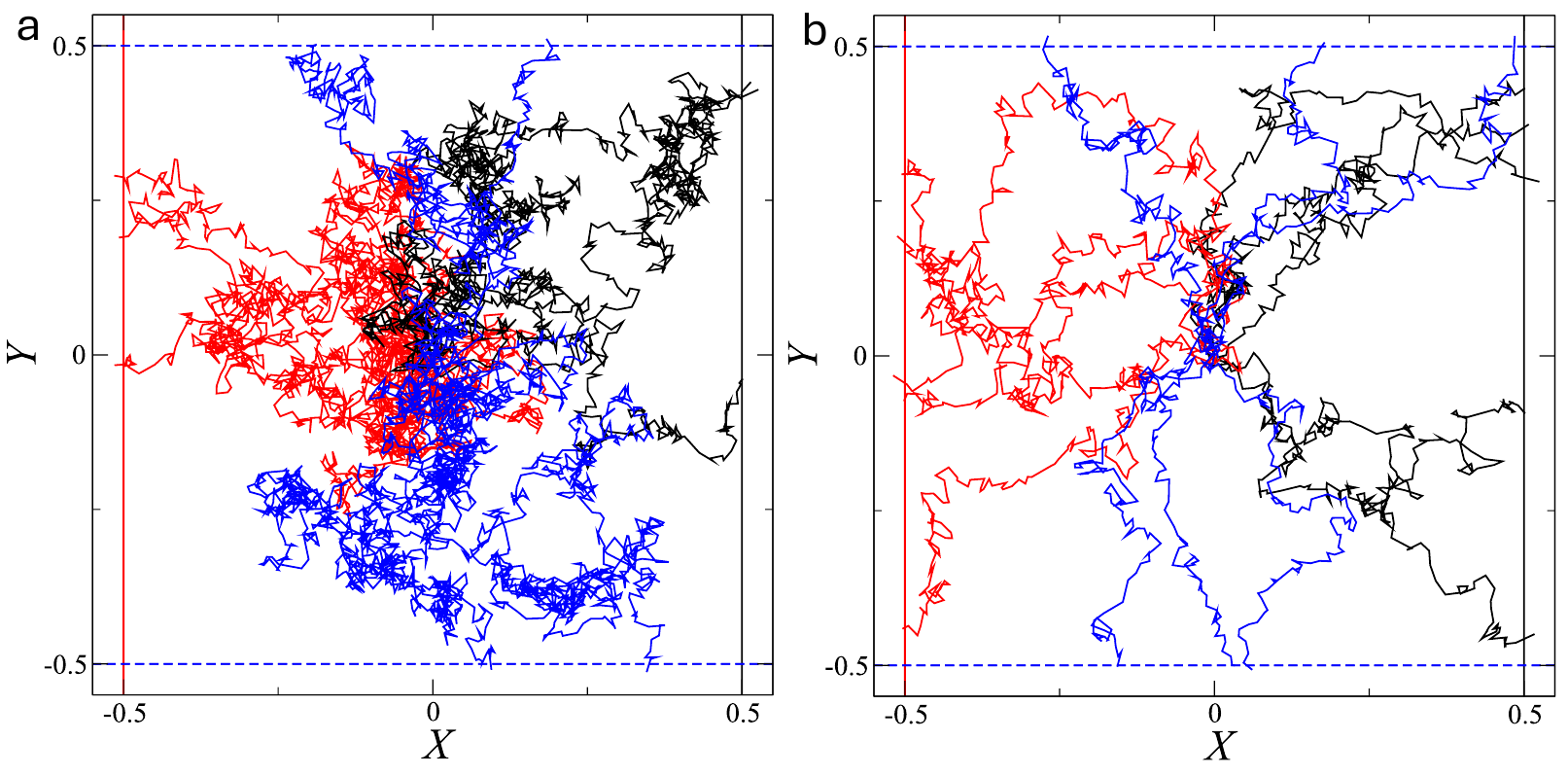}
    \caption{
    Representative planar trajectories
    $(X(\tau),Y(\tau))$ of an active Brownian particle
    subject to a constant bias directed along $+x$.
    Downhill trajectories reaching the absorbing boundary
    at $X=+1/2$ are shown in black, uphill trajectories
    reaching $X=-1/2$ in red, and selected trajectories
    with large transverse excursions in blue. No
    absorbing  boundary is imposed in the
    transverse direction. Panels (a) and (b) correspond
    to the dimensionless self-propulsion speeds $U=0.21$
    and $U=1.05$, respectively, at fixed dimensionless
    bias $F=9.66$. The planar trajectories are shown for
    illustration; all quantitative first passage
    calculations use the closed process $(X,\phi)$.
    }
    \label{fig_trajectories_2D}
\end{figure*}

Figure~\ref{fig_trajectories_2D} provides a planar
visualisation of representative trajectories generated
from the complete two-dimensional dynamics in
Eqs.~\eqref{eq:Langevin-dimless}. The trajectories are
classified according to their first arrival at the
vertical absorbing boundaries $X=\pm1/2$. Downhill
trajectories reaching $X=+1/2$ are shown in black,
whereas uphill trajectories reaching $X=-1/2$ are shown
in red. Selected trajectories with large transverse
excursions are highlighted in blue to illustrate the
broader planar motion of the particle. No absorbing boundary is imposed in the transverse
direction.
The comparison between the two panels illustrates how
increasing $U$ produces more persistent and
directionally correlated planar trajectories. The
transverse coordinate is displayed only for this
geometrical illustration and is not used in the
calculation of the projected first passage observables.

\begin{figure*}[t]
    \centering
    \includegraphics[width=0.8\textwidth]{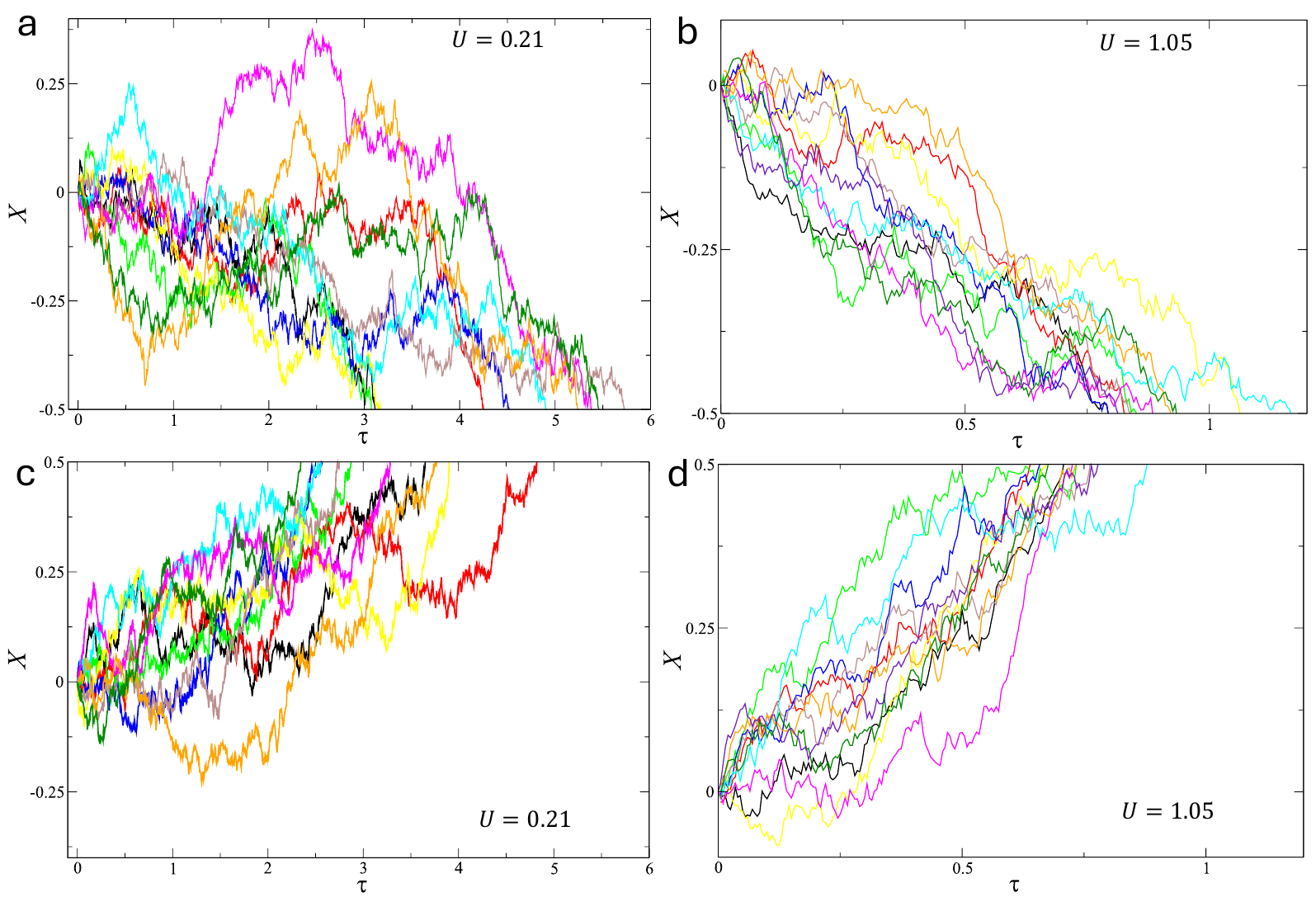}
    \caption{
    Representative projected trajectories $X(\tau)$
    generated from the closed process $(X,\phi)$ for an
    active Brownian particle undergoing first passage
    between the absorbing boundaries $X=\pm1/2$ in the
    presence of a constant bias directed along $+x$.
    Panels (a,b) show uphill trajectories terminating at
    $X=-1/2$, while panels (c,d) show downhill
    trajectories terminating at $X=+1/2$. The two
    columns correspond to the dimensionless
    self-propulsion parameters $U=0.21$
    [panels (a,c)] and $U=1.05$ [panels (b,d)], at fixed
    dimensionless bias $F=9.66$. Increasing persistence
    makes the projected motion more correlated and
    accentuates the contrast between uphill and downhill
    passages.
    }
    \label{fig_trajectories}
\end{figure*}

Figure~\ref{fig_trajectories} shows representative
projected trajectories $X(\tau)$ generated from the
closed process $(X,\phi)$. Panels (a,b) show uphill
trajectories terminating at $X=-1/2$, whereas panels
(c,d) show downhill trajectories terminating at
$X=+1/2$. Increasing $U$ makes the projected motion more
persistent and changes both the geometry and duration of
the crossings, while the external bias favours downhill
exits through $X=+1/2$. These examples anticipate the
quantitative results presented below: persistence modifies
the first passage time statistics, and the applied bias produces distinct uphill and downhill first passage behaviour.

The first passage observables considered below are the
direction-resolved first passage time probability
densities, the splitting probabilities for reaching
$X=+1/2$ or $X=-1/2$, and the associated moments of the
passage times. We next formulate these quantities at the
level of transition-path ensembles.

\section{Distribution of transition paths}
\label{sec:transition_paths}

We formulate the projected first passage problem in terms
of transition paths of the closed Markov process
$(X,\phi)$. Let $\sigma\in\{+1,-1\}$ denote the exit
direction, with $\sigma=+1$ corresponding to a downhill
passage through $X=+1/2$ and $\sigma=-1$ to an uphill
passage through $X=-1/2$. A transition path of duration
$\tau$ is a realisation
\begin{equation}
\omega
=
\{X(s),\phi(s)\}_{0\leq s\leq\tau},
\label{eq:transition_path}
\end{equation}
which starts at $X(0)=0$ and reaches the boundary
$X=\sigma/2$ for the first time at $s=\tau$:
\begin{equation}
X(\tau)=\frac{\sigma}{2},
\qquad
-\frac{1}{2}<X(s)<\frac{1}{2}
\quad\text{for }0<s<\tau .
\label{eq:first_passage_path}
\end{equation}
The initial orientation $\phi(0)=\phi_0$ is drawn
uniformly from $[0,2\pi)$, whereas the final orientation
is left unconstrained.

For a fixed initial orientation $\phi_0$, the probability
density of a path generated by
Eqs.~(\ref{eq:dimless_x}) and
(\ref{eq:dimless_phi}) can be written in
Onsager--Machlup form
\cite{OnsagerMachlup1953,Graham1977,DekerHaake1975}:
\begin{equation}
\mathcal P[\omega\mid\phi_0]
=
\mathcal N
\exp\!\left[-\mathcal A[\omega]\right],
\label{eq:path_probability}
\end{equation}
where $\mathcal N$ is a normalisation factor and
\begin{align}
\mathcal A[\omega]
&=
\frac{1}{4\widetilde D}
\int_0^\tau
\bigl[
\dot X(s)-\widetilde D F-U\cos\phi(s)
\bigr]^2\,ds
\nonumber\\
&\qquad
+
\frac{1}{4}
\int_0^\tau
\dot\phi(s)^2\,ds .
\label{eq:path_action}
\end{align}
Here and below, the overdot denotes differentiation with
respect to the dimensionless time $s$. Because the drift
in Eq.~(\ref{eq:dimless_x}) is independent of $X$, no
additional $X$-dependent Jacobian term appears in the
action.
Averaging over the uniform initial orientation gives the
path probability used below:
\begin{equation}
\mathcal P[\omega]
=
\frac{1}{2\pi}
\int_0^{2\pi}
d\phi_0\,
\mathcal P[\omega\mid\phi_0].
\label{eq:path_probability_averaged}
\end{equation}
The conditional density in the integrand is understood to
be supported on paths satisfying $\phi(0)=\phi_0$.

Let $\Omega_\sigma(\tau)$ denote the set of all first-passage paths
$\omega_\tau$ of duration $\tau$ that exit through $X=\sigma/2$.
The direction-resolved first passage time density is obtained by
summing the probability weights of all such paths, i.e. by the
functional integral
\begin{equation}
\rho_\sigma(\tau)
=
\int_{\Omega_\sigma(\tau)}
\mathcal D\omega\,
\mathcal P[\omega].
\label{eq:rho_path}
\end{equation}
Here $\mathcal D\omega$ denotes the path-integration measure over
the trajectories $\omega$.
Thus, $\rho_\sigma(\tau)\,d\tau$ is the joint
probability that the particle exits through
$X=\sigma/2$ and that the passage duration lies in
$[\tau,\tau+d\tau]$. Since the exit direction has not
been conditioned upon, $\rho_\sigma$ is not normalised
to one. Its integral gives the corresponding splitting
probability,
\begin{equation}
\Pi_\sigma
=
\int_0^\infty
\rho_\sigma(\tau)\,d\tau .
\label{eq:Pi_path}
\end{equation}
The first passage time probability density conditioned
on exit through $X=\sigma/2$ is therefore
\begin{equation}
p_\sigma(\tau)
=
\frac{\rho_\sigma(\tau)}{\Pi_\sigma},
\label{eq:conditional_fpt_pdf}
\end{equation}
and is normalised to one. The corresponding
direction-conditioned mean first passage time is
\begin{equation}
\langle\tau_\sigma\rangle
=
\int_0^\infty
\tau\,p_\sigma(\tau)\,d\tau.
\label{eq:conditional_mean_fpt}
\end{equation}

To compare the two directional path ensembles, we pair
each path with its spatially reflected counterpart:
\begin{equation}
(\mathcal R\omega)(s)
=
\bigl(-X(s),\phi(s)\bigr),
\qquad
0\leq s\leq\tau .
\label{eq:R_map}
\end{equation}
The reflection maps $\Omega_\sigma(\tau)$ bijectively onto
$\Omega_{-\sigma}(\tau)$. It preserves the duration,
the temporal ordering, and the complete orientation
history, while changing
$X(s)\to-X(s)$ and
$\dot X(s)\to-\dot X(s)$. It is therefore a spatial
reflection at fixed time, rather than a physical time
reversal. Applying it twice returns the original path,
$\mathcal R^2=1$.

We define the path-dependent asymmetry functional as
the logarithm of the probability ratio of a path and its
reflected partner:
\begin{equation}
\Sigma[\omega]
=
\ln
\frac{\mathcal P[\omega]}
{\mathcal P[\mathcal R\omega]}
=
\mathcal A[\mathcal R\omega]
-
\mathcal A[\omega].
\label{eq:Sigma_def}
\end{equation}
Log-ratios of probabilities assigned to paired paths are
widely used in stochastic thermodynamics and active
matter
\cite{Seifert2012,DabelowBoEichhorn2019}. In the present
problem, the pairing is defined by the spatial reflection
$\mathcal R$, not by time reversal.
Using Eq.~(\ref{eq:path_action}), we obtain
\begin{equation}
\Sigma[\omega]
=
F\bigl[X(\tau)-X(0)\bigr]
+
\frac{U}{\widetilde D}
\int_0^\tau
\cos\phi(s)\circ dX(s).
\label{eq:Sigma_explicit}
\end{equation}
The first term is fixed by the endpoints and represents
the contribution of the external bias. The second term
depends on the displacement increments and orientations
sampled throughout the passage and is therefore
trajectory dependent. The symbol $\circ dX$ denotes a
Stratonovich integral. In the present model, the
Stratonovich and It\^o forms coincide for this mixed term
because the translational and angular noises are
independent.

For a path in $\Omega_\sigma(\tau)$,
$X(\tau)-X(0)=\sigma/2$, and hence
\begin{equation}
\Sigma[\omega]
=
\frac{\sigma F}{2}
+
\frac{U}{\widetilde D}
\int_0^\tau
\cos\phi(s)\circ dX(s).
\label{eq:Sigma_directional}
\end{equation}
In the passive limit $U=0$, $\Sigma$ is constant within
each directional ensemble. Activity makes it
path-dependent because different passages sample
different orientational histories.
Since $\mathcal R^2=1$, it follows directly from the
definition of $\Sigma$ that
\begin{equation}
\Sigma[\mathcal R\omega]
=
-\Sigma[\omega].
\label{eq:Sigma_odd}
\end{equation}
Equation~(\ref{eq:Sigma_def}) also renders the corresponding
path-weight relation
\begin{equation}
\mathcal P[\mathcal R\omega]
=
e^{-\Sigma[\omega]}
\mathcal P[\omega].
\label{eq:path_weight_relation}
\end{equation}

We first use Eq.~(\ref{eq:path_weight_relation}) to
relate the uphill first passage time probability density to the downhill path ensemble. Integrating
it over the downhill ensemble gives
\begin{equation}
\int_{\Omega_+(\tau)}
\mathcal D\omega\,
\mathcal P[\mathcal R\omega]
=
\int_{\Omega_+(\tau)}
\mathcal D\omega\,
e^{-\Sigma[\omega]}
\mathcal P[\omega].
\label{eq:integrated_path_weight}
\end{equation}
The reflection pairs every downhill path of duration
$\tau$ with exactly one uphill path of the same duration,
and every uphill path is obtained in this way. Since the
reflection also preserves the path-integration measure,
the left-hand side of
Eq.~(\ref{eq:integrated_path_weight}) is the integral of
the path probability over the complete uphill ensemble:
\[
\int_{\Omega_+(\tau)}
\mathcal D\omega\,
\mathcal P[\mathcal R\omega]
=
\int_{\Omega_-(\tau)}
\mathcal D\omega\,
\mathcal P[\omega]
=
\rho_-(\tau).
\]
Equation~(\ref{eq:integrated_path_weight}) therefore
becomes
\begin{equation}
\rho_-(\tau)
=
\int_{\Omega_+(\tau)}
\mathcal D\omega\,
e^{-\Sigma[\omega]}
\mathcal P[\omega].
\label{eq:rho_minus_from_plus}
\end{equation}
Using
$p_-(\tau)=\rho_-(\tau)/\Pi_-$ then gives
\begin{equation}
p_-(\tau)
=
\frac{1}{\Pi_-}
\int_{\Omega_+(\tau)}
\mathcal D\omega\,
e^{-\Sigma[\omega]}
\mathcal P[\omega].
\label{eq:conditional_weighted_symmetry}
\end{equation}

For the symmetric representation, which is used in
Figs.~\ref{fig_WeightPDF}
and~\ref{fig:appendix_weighted_collapse}, we start from
Eq.~\eqref{eq:path_weight_relation}. Multiplying both
sides of this relation by
$e^{\Sigma[\omega]/2}$ gives
\[
e^{\Sigma[\omega]/2}
\mathcal P[\mathcal R\omega]
=
e^{-\Sigma[\omega]/2}
\mathcal P[\omega].
\]
Using the oddness relation
$\Sigma[\mathcal R\omega]
=-\Sigma[\omega]$ from
Eq.~(\ref{eq:Sigma_odd}), this becomes
\begin{equation}
e^{-\Sigma[\mathcal R\omega]/2}
\mathcal P[\mathcal R\omega]
=
e^{-\Sigma[\omega]/2}
\mathcal P[\omega].
\label{eq:half_weighted_pair}
\end{equation}
Thus, a path and its reflected partner carry the same
half-weighted probability.

We now integrate
Eq.~(\ref{eq:half_weighted_pair}) over all downhill paths
of duration $\tau$. Using the same bijective,
measure-preserving reflection argument as in the
derivation of Eq.~(\ref{eq:rho_minus_from_plus}), the
left-hand side becomes the corresponding integral over
the complete uphill ensemble. Hence
\begin{equation}
\int_{\Omega_-(\tau)}
\mathcal D\omega\,
e^{-\Sigma[\omega]/2}
\mathcal P[\omega]
=
\int_{\Omega_+(\tau)}
\mathcal D\omega\,
e^{-\Sigma[\omega]/2}
\mathcal P[\omega].
\label{eq:half_weighted_fixed_duration}
\end{equation}
The unnormalised half-weighted first passage time densities are
therefore equal at every duration.

Integrating Eq.~(\ref{eq:half_weighted_fixed_duration})
over all passage times shows that the two directional
normalisations are also equal. We denote their common
value by
\begin{equation}
\begin{aligned}
\mathcal Z^{(1/2)}
&\equiv
\int_0^\infty d\tau'
\int_{\Omega_+(\tau')}
\mathcal D\omega\,
e^{-\Sigma[\omega]/2}
\mathcal P[\omega]
\\
&=
\int_0^\infty d\tau'
\int_{\Omega_-(\tau')}
\mathcal D\omega\,
e^{-\Sigma[\omega]/2}
\mathcal P[\omega].
\end{aligned}
\label{eq:half_weighted_normalisation}
\end{equation}
The separately normalised half-weighted first passage time densities are then
\begin{equation}
p_\sigma^{(1/2)}(\tau)
=
\frac{1}{\mathcal Z^{(1/2)}}
\int_{\Omega_\sigma(\tau)}
\mathcal D\omega\,
e^{-\Sigma[\omega]/2}
\mathcal P[\omega],
\qquad
\sigma=\pm1.
\label{eq:half_weighted_density}
\end{equation}
Equation~(\ref{eq:half_weighted_fixed_duration})
therefore renders
\begin{equation}
p_+^{(1/2)}(\tau)
=
p_-^{(1/2)}(\tau).
\label{eq:half_weighted_symmetry}
\end{equation}

The one-sided and symmetric relations follow from the
same reflected-path probability ratio. In
Eq.~(\ref{eq:conditional_weighted_symmetry}), the full
factor $e^{-\Sigma}$ converts the downhill path weights
into the corresponding uphill path weights. In
Eq.~(\ref{eq:half_weighted_symmetry}), the factor is
divided equally between each path and its reflected
partner. In the passive limit, $U=0$, one has
$\Sigma=\sigma F/2$ within each directional ensemble.
The resulting constant factors cancel when the
directional densities are normalised, so that
$p_\sigma^{(1/2)}(\tau)=p_\sigma(\tau)$. The
half-weighted equality then reduces to
\begin{equation}
p_+(\tau)=p_-(\tau).
\label{eq:passive_conditional_symmetry}
\end{equation}

\section{Fokker--Planck formulation and asymptotic regimes}
\label{sec:survival_splitting}

The purpose of this section is to determine the
direction-conditioned mean first passage time
$\langle\tau_\sigma\rangle$ from the splitting
probability and the corresponding first directional
moment, and to analyse its behaviour in the small- and
large-$U$ regimes. We first define these quantities for a
specified initial state and formulate the associated
forward and backward Fokker--Planck problems. The detailed perturbative calculations and some  numerical tests are presented in Appendix~C.

For the dimensionless Langevin equations
\eqref{eq:dimless_x} and \eqref{eq:dimless_phi}, let
$G(X,\phi,\tau\mid X_0,\phi_0)$ denote the conditional
probability density for the particle to be at
$(X,\phi)$ at time $\tau$, given the initial state
$(X_0,\phi_0)$. It satisfies the forward
Fokker--Planck equation
\begin{equation}
\frac{\partial G}{\partial\tau}
=
-\frac{\partial}{\partial X}
\left[
\left(\widetilde D F+U\cos\phi\right)G
\right]
+
\widetilde D\,\frac{\partial^2G}{\partial X^2}
+
\frac{\partial^2G}{\partial\phi^2},
\label{eq:FP_forward}
\end{equation}
with initial condition
\begin{equation}
G(X,\phi,0\mid X_0,\phi_0)
=
\delta(X-X_0)\,
\delta_{2\pi}(\phi-\phi_0),
\label{eq:FP_initial}
\end{equation}
where $\delta_{2\pi}$ denotes the periodic delta function
in the angular variable. The absorbing boundary
conditions at $X=\pm1/2$ give
\begin{equation}
G\!\left(
-\frac{1}{2},\phi,\tau\mid X_0,\phi_0
\right)
=
G\!\left(
\frac{1}{2},\phi,\tau\mid X_0,\phi_0
\right)
=
0,
\label{eq:absorbing_bc}
\end{equation}
for $0\leq\phi<2\pi$ and $\tau>0$, together with
$2\pi$-periodicity in $\phi$.

For a specified initial state $(X_0,\phi_0)$, the direction-resolved
first passage time probability density $\rho_\sigma(\tau\mid X_0,\phi_0)$ is given by the outward flux 
of the translational probability current through the
boundary $X=\sigma/2$:
\begin{equation}
\rho_\sigma(\tau\mid X_0,\phi_0)
=
\sigma
\int_0^{2\pi}d\phi\,
\left.
J(X,\phi,\tau\mid X_0,\phi_0)
\right|_{X=\sigma/2}.
\label{eq:rho_sigma_flux}
\end{equation}
The translational probability current associated with
Eq.~\eqref{eq:FP_forward} is
\begin{equation}
J(X,\phi,\tau\mid X_0,\phi_0)
=
\left(\widetilde D F+U\cos\phi\right)G
-
\widetilde D\,\frac{\partial G}{\partial X}.
\label{eq:current}
\end{equation}
The factor $\sigma$ in Eq.~\eqref{eq:rho_sigma_flux}
is the outward unit normal at the boundary
$X=\sigma/2$, so that $\sigma J$ is the component of
the probability current projected onto the outward
normal.

$\rho_\sigma(\tau\mid X_0,\phi_0)$ extends the
direction-resolved density $\rho_\sigma(\tau)$ introduced
in Sec.~\ref{sec:transition_paths} to a specified initial
position and orientation; recall that $\rho_\sigma(\tau)$  corresponds to
$X_0=0$ and an isotropically distributed initial
orientation.
The zeroth and first time moments of
$\rho_\sigma(\tau\mid X_0,\phi_0)$ are defined
\begin{align}
\Pi_\sigma(X_0,\phi_0)
&=
\int_0^\infty
\rho_\sigma(\tau\mid X_0,\phi_0)\,d\tau,
\label{eq:splitting_def}\\
M_\sigma(X_0,\phi_0)
&=
\int_0^\infty
\tau\,\rho_\sigma(\tau\mid X_0,\phi_0)\,d\tau,
\label{eq:first_directional_moment}
\end{align}
respectively, where $\Pi_\sigma(X_0,\phi_0)$ describes the splitting 
probability for exit through $X=\sigma/2$.
%
For an isotropically distributed $\phi_0$, we
define
\begin{align}
\Pi_\sigma(X_0)
&=
\frac{1}{2\pi}
\int_0^{2\pi}
\Pi_\sigma(X_0,\phi_0)\,d\phi_0,
\label{eq:Pi_orientation_average}\\
M_\sigma(X_0)
&=
\frac{1}{2\pi}
\int_0^{2\pi}
M_\sigma(X_0,\phi_0)\,d\phi_0,
\label{eq:M_orientation_average}
\end{align}
such that for the centred isotropic initial ensemble considered in
Sec.~\ref{sec:transition_paths},
$\Pi_\sigma\equiv\Pi_\sigma(0)$ and
$M_\sigma\equiv M_\sigma(0)$. Using
Eq.~\eqref{eq:conditional_mean_fpt}, the
direction-conditioned mean first passage time can then
be written as
\begin{equation}
\langle\tau_\sigma\rangle
=
\frac{M_\sigma}{\Pi_\sigma}.
\label{eq:conditional_mean_from_moments}
\end{equation}

The splitting probability and first directional
moment can be determined without solving the
time-dependent forward problem. They satisfy backward
Kolmogorov equations
\cite{Grebenkov2015FirstExit,Lefebvre1989FirstPassage}
generated by the operator
\begin{equation}
\mathcal L^\dagger
=
\left(\widetilde D F+U\cos\phi_0\right)
\frac{\partial}{\partial X_0}
+
\widetilde D\,
\frac{\partial^2}{\partial X_0^2}
+
\frac{\partial^2}{\partial\phi_0^2},
\label{eq:backward_operator}
\end{equation}
which acts on the initial coordinates $(X_0,\phi_0)$.
For instance, $\Pi_\sigma(X_0,\phi_0)$ obeys 
\begin{equation}
\mathcal L^\dagger
\Pi_\sigma(X_0,\phi_0)
=
0,
\label{eq:splitting_backward}
\end{equation}
with boundary conditions
\begin{equation}
\Pi_\sigma\!\left(
\frac{\sigma}{2},\phi_0
\right)
=
1,
\qquad
\Pi_\sigma\!\left(
-\frac{\sigma}{2},\phi_0
\right)
=
0,
\label{eq:Pi_sigma_bc}
\end{equation}
and $2\pi$-periodicity in $\phi_0$. Since the two exit events are complementary, it is
sufficient to solve for one of them.
In turn, $M_\sigma(X_0,\phi_0)$ can be obtained from the following backward equation
\begin{equation}
\mathcal L^\dagger
M_\sigma(X_0,\phi_0)
=
-\Pi_\sigma(X_0,\phi_0),
\label{eq:directional_moment_backward}
\end{equation}
with homogeneous boundary conditions
\begin{equation}
M_\sigma\!\left(
\pm\frac{\sigma}{2},\phi_0
\right)
=
0.
\label{eq:directional_moment_bc}
\end{equation}


We now use the backward formulation (\ref{eq:splitting_backward})--(\ref{eq:directional_moment_bc})
to derive perturbatively the direction-conditioned mean first passage time
$\langle\tau_\sigma\rangle$ \eqref{eq:conditional_mean_from_moments}.
The corresponding perturbation hierarchy is presented in more detail in 
Appendix~\ref{app:weak_persistence}.

\subsection{Small $U$ expansion and rapid angular relaxation asymptotics}
\label{sec:5A}

Two sequential asymptotic steps are involved. First, the solutions of the
backward problems are expanded in the weak-persistence parameter
$U=\ell_p/L\ll1$ at fixed $\widetilde D$ and $F$. Second, the resulting
coefficients are evaluated in the limit of rapid angular relaxation. We
introduce a small parameter 
\[
\varepsilon
\equiv
\widetilde D.
\]
In the angular Fourier expansions of $\Pi_\sigma(X_0,\phi_0)$ and $M_\sigma(X_0,\phi_0)$, the $n=1$ angular modes, represented here by terms
proportional to $\cos\phi_0$, relax under the dimensionless angular
operator $\partial_{\phi_0}^2$ with rate $1$. For spatial variations
on the scale of the interval, the translational diffusion and
force-induced drift terms are instead of orders $\varepsilon$ and
$\varepsilon F$, respectively. Thus, at fixed $F$, rapid angular
relaxation corresponds to $\varepsilon\ll1$. The expansion in $U$ is performed first at fixed
$\varepsilon$, and the limit $\varepsilon\to0$ is subsequently applied
to its coefficients. No scaling relation between $U$ and $\varepsilon$
is imposed in deriving these coefficients.

Before expanding the backward solutions in $U$, we note
that isotropic averaging over the initial orientation makes
the resulting first-passage observables even functions of
$U$. Indeed, under $U\to-U$ the shift
$\phi\to\phi+\pi$ leaves the projected active drift
unchanged, since
$(-U)\cos(\phi+\pi)=U\cos\phi$.
The angular dynamics, the boundary conditions, and the
uniform distribution of the initial orientation are also
unchanged by this shift. Consequently, the orientationally
averaged splitting probability and first directional moment
are both even functions of $U$, and so is their ratio
$\langle\tau_\sigma\rangle$. At fixed $\varepsilon>0$
the backward problems depend regularly on $U$, and hence
\begin{equation}
\langle\tau_\sigma\rangle
=
T_0(\varepsilon,F)
+
U^2T_\sigma^{(2)}(\varepsilon,F)
+
O(U^4).
\label{eq:conditional_time_U_expansion}
\end{equation}

\paragraph {Common leading asymptotics of the direction-conditioned mean first passage times.}
For an isotropic distribution of initial orientations, $X_0=0$, and
fixed $F$, the $O(U^2)$ coefficients of
$\langle\tau_+\rangle$ and $\langle\tau_-\rangle$ have the same
leading asymptotic behaviour for $\varepsilon\to0$:
\begin{equation}
T_\sigma^{(2)}(\varepsilon,F)
=
-\frac{1}{16\varepsilon^2}
\operatorname{sech}^{2}\!\left(\frac{F}{4}\right)
+
R_\sigma(\varepsilon,F),
\label{eq:asymptotic_identity_T2}
\end{equation}
where
\begin{equation}
R_\sigma(\varepsilon,F)
=
O(\varepsilon^{-3/2}),
\qquad
\varepsilon\to0.
\label{eq:T2_remainder_scaling}
\end{equation}
Thus $R_\sigma$ is subleading relative to the common
$O(\varepsilon^{-2})$ term. The estimate
$R_\sigma=O(\varepsilon^{-3/2})$ applies to each exit
direction separately and does not determine the leading
order of their difference: contributions of order
$O(\varepsilon^{-3/2})$ may cancel in
$R_+-R_-$. The directional difference is therefore of
relative order $O(\sqrt{\varepsilon})$ or smaller, and in
particular
\begin{equation}
\lim_{\varepsilon\to0}
\varepsilon^2
\left[
T_+^{(2)}(\varepsilon,F)
-
T_-^{(2)}(\varepsilon,F)
\right]
=
0.
\label{eq:asymptotic_directional_identity}
\end{equation}
The derivation from the backward hierarchy is given in
Appendix~\ref{app:weak_persistence}, while the convergence of the rescaled
coefficients $\varepsilon^2 T_{\sigma}^{(2)}$ to the common limiting
value in Eq.~\eqref{eq:asymptotic_identity_T2} is verified numerically
in Fig.~\ref{fig:T2_validation} of
Appendix~\ref{app:weak_persistence}.

The passive contribution in
Eq.~(\ref{eq:conditional_time_U_expansion}) reads
\begin{equation}
T_0(\varepsilon,F)
=
\frac{1}{2\varepsilon F}
\tanh\!\left(\frac{F}{4}\right),
\qquad
F\neq0,
\label{eq:passive_conditional_time}
\end{equation}
with the continuous limit
$T_0(\varepsilon,F\rightarrow 0)=1/(8\varepsilon)$.
Combining Eqs.~(\ref{eq:conditional_time_U_expansion}) and
(\ref{eq:asymptotic_identity_T2}) gives
\begin{align}
\langle\tau_\sigma\rangle
&=
\frac{1}{2\varepsilon F}
\tanh\!\left(\frac{F}{4}\right)
-
\frac{U^2}{16\varepsilon^2}
\operatorname{sech}^{2}\!\left(\frac{F}{4}\right)
\nonumber\\
&\quad
+
U^2R_\sigma(\varepsilon,F)
+
O(U^4).
\label{eq:conditional_time_asymptotic}
\end{align}
The passive term and the explicitly displayed $O(U^2)$
term are common to the two passage directions. Any
directional dependence at $O(U^2)$ is contained in
$R_\sigma$. Its subleading character refers only to the
subsequent $\varepsilon\to0$ asymptotics: $R_\sigma$
remains part of the $O(U^2)$ coefficient of the
weak-persistence expansion.

The common leading $O(U^2)$ term in
Eq.~\eqref{eq:conditional_time_asymptotic} has a simple
effective-diffusion interpretation. It coincides with the
$O(U^2)$ term of the passive drift-diffusion result for the mean first passage time when
the translational diffusivity is replaced by
\begin{equation}
D_{\rm eff}
=
\varepsilon+\frac{U^2}{2},
\label{eq:Deff_wp}
\end{equation}
while the drift $v=\varepsilon F$ is held fixed.
A one-dimensional drift-diffusion process with this
effective diffusivity is
\begin{equation}
dX
=
\varepsilon F\,d\tau
+
\sqrt{2D_{\rm eff}}\,dW_\tau,
\label{eq:weak_persistence_SDE}
\end{equation}
where $W_\tau$ is a standard one-dimensional Wiener
process.
For this effective process, the passive finite-interval
result implies that the uphill and downhill conditional
first passage (from the interval
centre) time probability densities are identical
\cite{dagbug:2009}, and a common direction-conditioned
mean first passage time is
\begin{equation}
\langle \tau \rangle_{\mathrm{eff}}
=
\frac{1}{2\varepsilon F}
\tanh\!\left(
\frac{\varepsilon F}{4D_{\mathrm{eff}}}
\right),
\label{eq:wp_conditional_mean}
\end{equation}
with the unbiased limit for $F\to0$
\begin{equation}
\langle \tau \rangle_{\mathrm{eff}}
=
\frac{1}{8D_{\mathrm{eff}}}.
\end{equation}
Expanding Eq.~\eqref{eq:wp_conditional_mean} through
$O(U^2)$ while keeping the drift coefficient
$\varepsilon F$ fixed renders
\begin{align}
\frac{1}{2\varepsilon F}
\tanh\!\left(
\frac{\varepsilon F}{4D_{\rm eff}}
\right)
&=
\frac{1}{2\varepsilon F}
\tanh\!\left(\frac{F}{4}\right)
\nonumber\\
&\quad
-
\frac{U^2}{16\varepsilon^2}
\operatorname{sech}^{2}\!\left(\frac{F}{4}\right)
+
O(U^4),
\label{eq:effective_time_expansion}
\end{align}
which reproduces the passive contribution and the common
leading $\varepsilon^{-2}$ part of the $O(U^2)$ correction
in Eq.~\eqref{eq:conditional_time_asymptotic}.

The remainder $R_\sigma$ originates from the non-uniform
behaviour of the expansion near the absorbing boundaries
in the limit of rapid angular relaxation.
As shown in
Appendix~\ref{app:weak_persistence}, the $O(U)$
coefficients $\pi_{1,\sigma}$ and $m_{1,\sigma}$ of the
$\cos\phi_0$ terms in the splitting probability and first
directional moment approach, away from the endpoints,
the spatial derivatives
$\pi'_{0,\sigma}\equiv\partial_{X_0}\pi_{0,\sigma}$ and
$m'_{0,\sigma}\equiv\partial_{X_0}m_{0,\sigma}$ of the
corresponding passive coefficients.
These interior approximations do not satisfy the
boundary conditions because
$\pi'_{0,\sigma}$ and $m'_{0,\sigma}$ are nonzero at
$X_0=\pm1/2$, whereas
$\pi_{1,\sigma}$ and $m_{1,\sigma}$ must vanish there.
Boundary-layer corrections are therefore required near
the two absorbing boundaries.
Spatial derivatives of these corrections
enter the equations for the $O(U^2)$ coefficients
$\pi_{20,\sigma}$ and $m_{20,\sigma}$, producing
corresponding contributions to both
coefficients. When $\pi_{20,\sigma}$ and
$m_{20,\sigma}$ are combined to obtain
$T_\sigma^{(2)}$, these boundary-layer contributions give rise to the remainder
$R_\sigma(\varepsilon,F)$. The direction-dependent part
of this remainder produces the difference between the
uphill and downhill $O(U^2)$ corrections. Since the effective
drift-diffusion description contains no counterpart of
the endpoint corrections required by the full
$(X_0,\phi_0)$ backward problem, it reproduces the common
leading term but not $R_\sigma$.

The same effective drift-diffusion process gives an
approximation to the splitting probabilities. For an
initial position $X_0$, the probability of reaching the
right boundary before the left one is
\begin{equation}
\Pi_+^{\mathrm{eff}}(X_0)
=
\frac{
1-\exp\!\left[
-\varepsilon F(X_0+1/2)/D_{\mathrm{eff}}
\right]
}{
1-\exp\!\left[
-\varepsilon F/D_{\mathrm{eff}}
\right]
}.
\end{equation}
For the symmetric initial condition $X_0=0$, this gives
\begin{align}
\Pi_+^{\rm eff}
&=
\frac{1}
{1+\exp[-\varepsilon F/(2D_{\rm eff})]},
\nonumber\\
\Pi_-^{\rm eff}
&=
1-\Pi_+^{\mathrm{ eff}}.
\label{eq:W_wp}
\end{align}
Expanding
Eqs.~\eqref{eq:W_wp} in powers of $U$ reproduces the
$O(U^2)$ contribution obtained from the weak-persistence
expansion after retaining its leading term as
$\varepsilon\to0$ at fixed $F$.


\subsection{Strong-persistence limit: frozen-orientation
estimate of the mean first passage times}

Here, we consider the strong-persistence regime
\begin{equation}
U=\frac{\ell_p}{L}\gg1.
\end{equation}
In the dimensionless variables, the angular relaxation
time is of order unity. For $\tau\ll1$, the angular displacement accumulated
during the passage is small, and the orientation can be
held fixed at its initial value:
\begin{equation}
\phi(s)\simeq\phi_0,
\qquad 0\leq s\leq\tau .
\label{eq:frozen_orientation}
\end{equation}
For each fixed $\phi_0$, the projected motion therefore
reduces to a one-dimensional constant drift diffusion problem,
\begin{equation}
dX
=
v(\phi_0)\,d\tau
+
\sqrt{2\widetilde D}\,dW_\tau,
\label{eq:frozen_orientation_dynamics}
\end{equation}
where $v(\phi_0)=\widetilde D F+U\cos\phi_0$. The approximation requires the first passage time to be
short compared with the angular relaxation time. This condition may fail
for initial orientations near the angles at which
$v(\phi_0)=0$, because the projected drift then becomes
small and the crossing time is no longer reduced by
increasing $U$.
At $X_0=0$, this drift-diffusion problem has the same
direction-conditioned mean first passage time for the two
absorbing boundaries~\cite{dagbug:2009}
\begin{equation}
\left\langle\tau_+\right\rangle_{\phi_0}
=
\left\langle\tau_-\right\rangle_{\phi_0}
\simeq
\frac{1}{2v(\phi_0)}
\tanh\!\left[
\frac{v(\phi_0)}{4\widetilde D}
\right],
\label{eq:frozen_orientation_conditional_mean}
\end{equation}
with the splitting
probability at fixed $\phi_0$ given by
\begin{equation}
\Pi_\sigma(0,\phi_0)
\simeq
\frac{1}{
1+\exp\!\left[
-\sigma v(\phi_0)/(2\widetilde D)
\right]
}.
\label{eq:frozen_orientation_splitting}
\end{equation}
Although the initial orientation is uniformly distributed,
the trajectories that exit through $X=\sigma/2$ do not
sample all values of $\phi_0$ with equal probability.
The contribution of a given initial orientation to the
exit-$\sigma$ ensemble is proportional to
$\Pi_\sigma(0,\phi_0)$.
The first directional moment at fixed $\phi_0$ can be written as 
\begin{align}
M_\sigma(0,\phi_0)
&=
\Pi_\sigma(0,\phi_0)
\left\langle\tau_\sigma\right\rangle_{\phi_0}
\nonumber\\
&\simeq
\Pi_\sigma(0,\phi_0)
\frac{1}{2v(\phi_0)}
\tanh\!\left[
\frac{v(\phi_0)}{4\widetilde D}
\right],
\label{eq:frozen_orientation_first_moment}
\end{align}
where we used \eqref{eq:frozen_orientation_conditional_mean} in the second line. 
Substituting \eqref{eq:frozen_orientation_first_moment}
into the orientational average defining $M_\sigma$ gives
\begin{equation}
M_\sigma
\simeq
\frac{1}{2\pi}
\int_0^{2\pi}d\phi_0\,
\Pi_\sigma(0,\phi_0)
\frac{1}{2v(\phi_0)}
\tanh\!\left[
\frac{v(\phi_0)}{4\widetilde D}
\right].
\label{eq:frozen_orientation_averaged_moment}
\end{equation}
Using
$\langle\tau_\sigma\rangle=M_\sigma/\Pi_\sigma$,
with $M_\sigma$ in \eqref{eq:frozen_orientation_averaged_moment}
and $\Pi_\sigma$ obtained by averaging
$\Pi_\sigma(0,\phi_0)$ in \eqref{eq:frozen_orientation_splitting} over the initial orientation, gives
\begin{equation}
\langle\tau_\sigma\rangle
\simeq
\frac{
\displaystyle
\int_0^{2\pi}d\phi_0\,
\Pi_\sigma(0,\phi_0)
\frac{1}{2v(\phi_0)}
\tanh\!\left[
\frac{v(\phi_0)}{4\widetilde D}
\right]
}{
\displaystyle
\int_0^{2\pi}d\phi_0\,
\Pi_\sigma(0,\phi_0)
}.
\label{eq:frozen_orientation_averaged_mean}
\end{equation}
For each fixed initial orientation $\phi_0$, the
direction-conditioned mean first passage times to the two
boundaries are equal. After averaging over $\phi_0$, $\langle\tau_+\rangle$ and
$\langle\tau_-\rangle$ can differ because the two exit
ensembles assign different weight
$\Pi_{\sigma}(0,\phi_0)$ to the same set of initial
orientations.

\subsection{Implications for the numerical results}
\label{sec:5C}

The weak-persistence and frozen-orientation limits explain
different features of the numerical results in
Sec.~\ref{sec:numerics}. In the weak-persistence regime, the common leading
$O(U^2)$ correction to the direction-conditioned mean
first passage times is reproduced by the passive
drift-diffusion result with
$D_{\rm eff}=\widetilde D+\frac{U^2}{2}$.
For the centred initial condition, this leading correction is the
same for the two exits, while any directional difference is
contained in the boundary-sensitive remainders $R_+$ and
$R_-$.

As $U$ approaches or exceeds unity, the dependence on the
initial orientation must be retained. 
In the frozen-orientation approximation, the projected
drift $v(\phi_0)$ and the exit-dependent weights
$\Pi_\sigma(0,\phi_0)$ retain the dependence on the initial
orientation. This exit-dependent weighting produces the separation between
$\langle\tau_+\rangle$ and $\langle\tau_-\rangle$, and the
asymmetry of the unweighted directional first passage
statistics observed numerically.

These asymptotic results concern the unweighted quantities
$\Pi_\sigma$, $\rho_\sigma(\tau)$, and
$\langle\tau_\sigma\rangle$. The reflection-based
path-reweighting relation derived in
Sec.~\ref{sec:transition_paths} is exact and applies
independently of the persistence regime. Thus, the
unweighted uphill and downhill statistics may differ
strongly, while the corresponding normalised half-weighted first passage time distributions remain equal.

\begin{figure*}[t]
\includegraphics[width=0.8\textwidth]{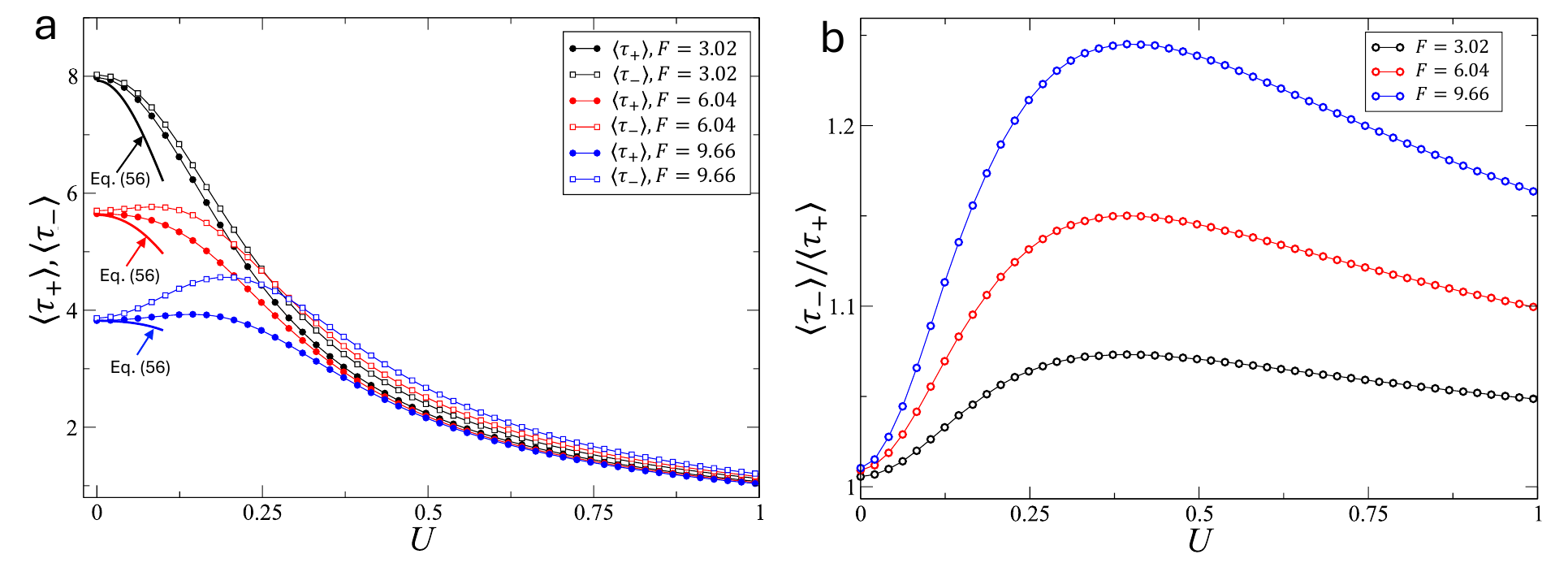}
\caption{(a) Mean first passage times for downhill and uphill crossings, \(\langle \tau_+\rangle\) and \(\langle \tau_-\rangle\), plotted as functions of the persistence parameter \(U\) for several values of the dimensionless bias \(F\), at fixed \(\widetilde D = 0.013\). The solid curves show the effective drift-diffusion prediction of Eq.~\eqref{eq:wp_conditional_mean}, whose small $U$ expansion reproduces the common leading $O(U^2)$ asymptotic contribution derived in Sec.~\ref{sec:5A}. The boundary-sensitive remainder \(R_\sigma\) is not included in these curves; any directional separation at \(O(U^2)\) is contained in the difference \(R_- - R_+\). Panel~(b) shows that the ratio
$\langle\tau_-\rangle/\langle\tau_+\rangle$
depends non-monotonically on the persistence parameter.
For the values of $F$ shown, its maximum increases and
shifts to larger values of $U$ as the external bias is
increased.
}
\label{fig_mean_times}
\end{figure*}

\begin{figure*}[htb]
\includegraphics[width=0.8\textwidth]{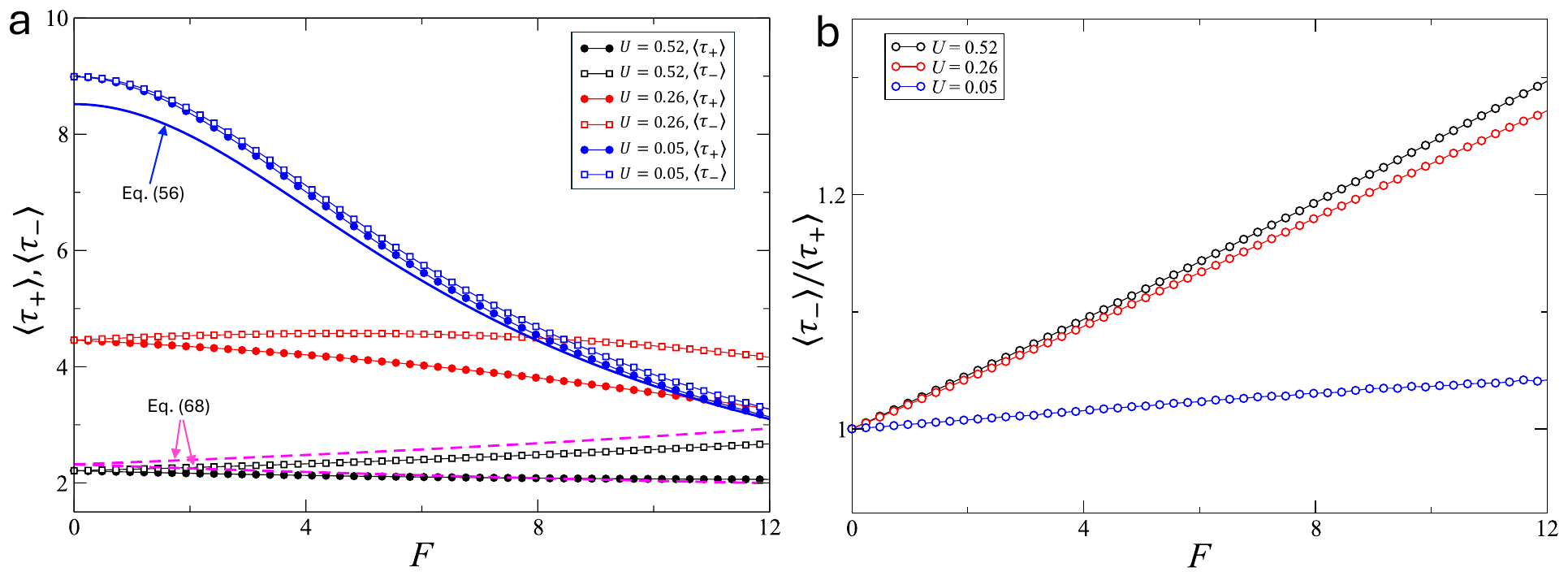}
\caption{(a) Mean first passage times for downhill and uphill crossings, \(\langle \tau_+\rangle\) and \(\langle \tau_-\rangle\), shown as functions of the dimensionless bias \(F\) for several values of the persistence parameter \(U\), at fixed \(\widetilde D = 0.013\). The solid curve labelled Eq.~\eqref{eq:wp_conditional_mean} shows the effective drift-diffusion results.
For $U=0.52$, the dashed magenta curves show the
frozen-orientation prediction in Eq.~\eqref{eq:frozen_orientation_averaged_mean}, where the upper and lower curves represent  \(\langle \tau_-\rangle\) and \(\langle \tau_+\rangle\), respectively.
(b) Corresponding ratio \(\langle \tau_-\rangle/\langle \tau_+\rangle\) as a function of \(F\). The figure shows that increasing the external bias
increases the difference between the downhill and uphill
mean first passage times, with a more pronounced
separation at larger 
$U$.}
\label{fig_mean_times_vsf}
\end{figure*}

\begin{figure*}[htb]
\includegraphics[width=0.8\textwidth]{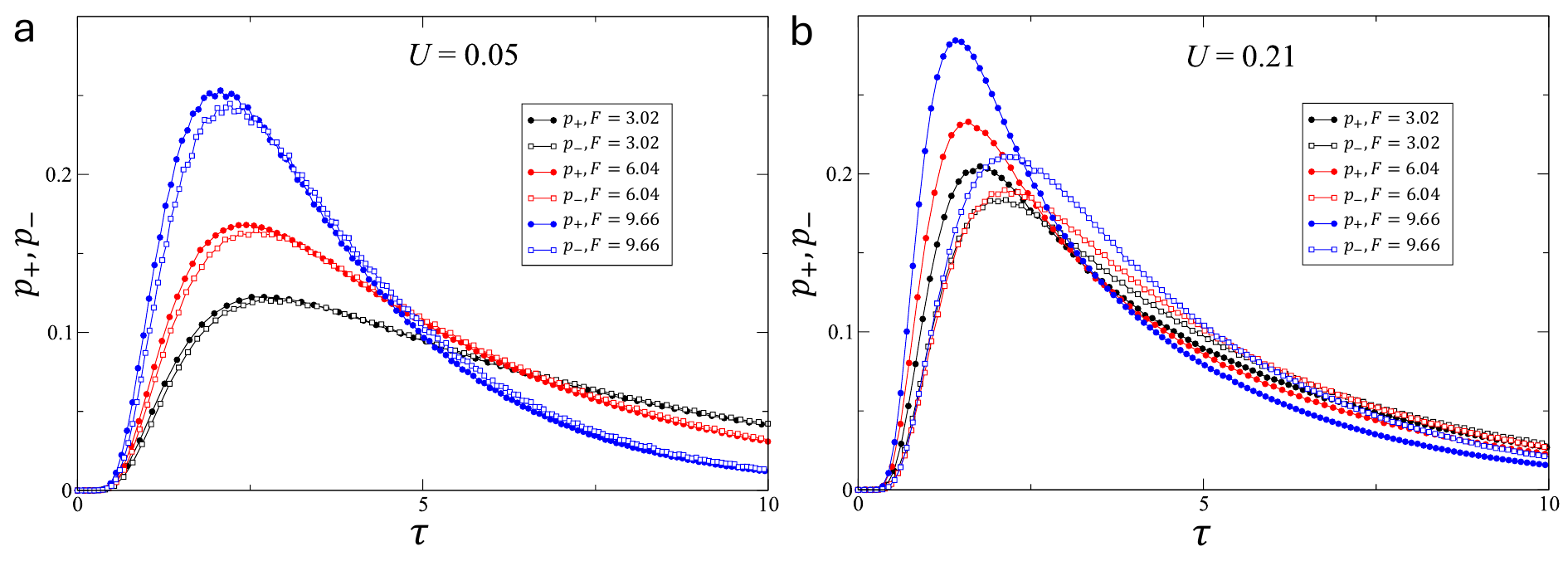}
\caption{Conditional first passage time probability densities for
downhill (filled symbols) and uphill (open symbols) passages, shown for (a) $U=0.05$ and (b) $U=0.21$. Different curves correspond to different values of the dimensionless bias $F$, at fixed $\widetilde D=0.013$. The comparison illustrates how increasing persistence enhances the asymmetry between uphill and downhill first passage time distributions, shifting the downhill peak to shorter times and broadening the uphill tail.}
\label{fig_PDFs_example}
\end{figure*}

\begin{figure}[htb]
\includegraphics[width=0.45\textwidth]{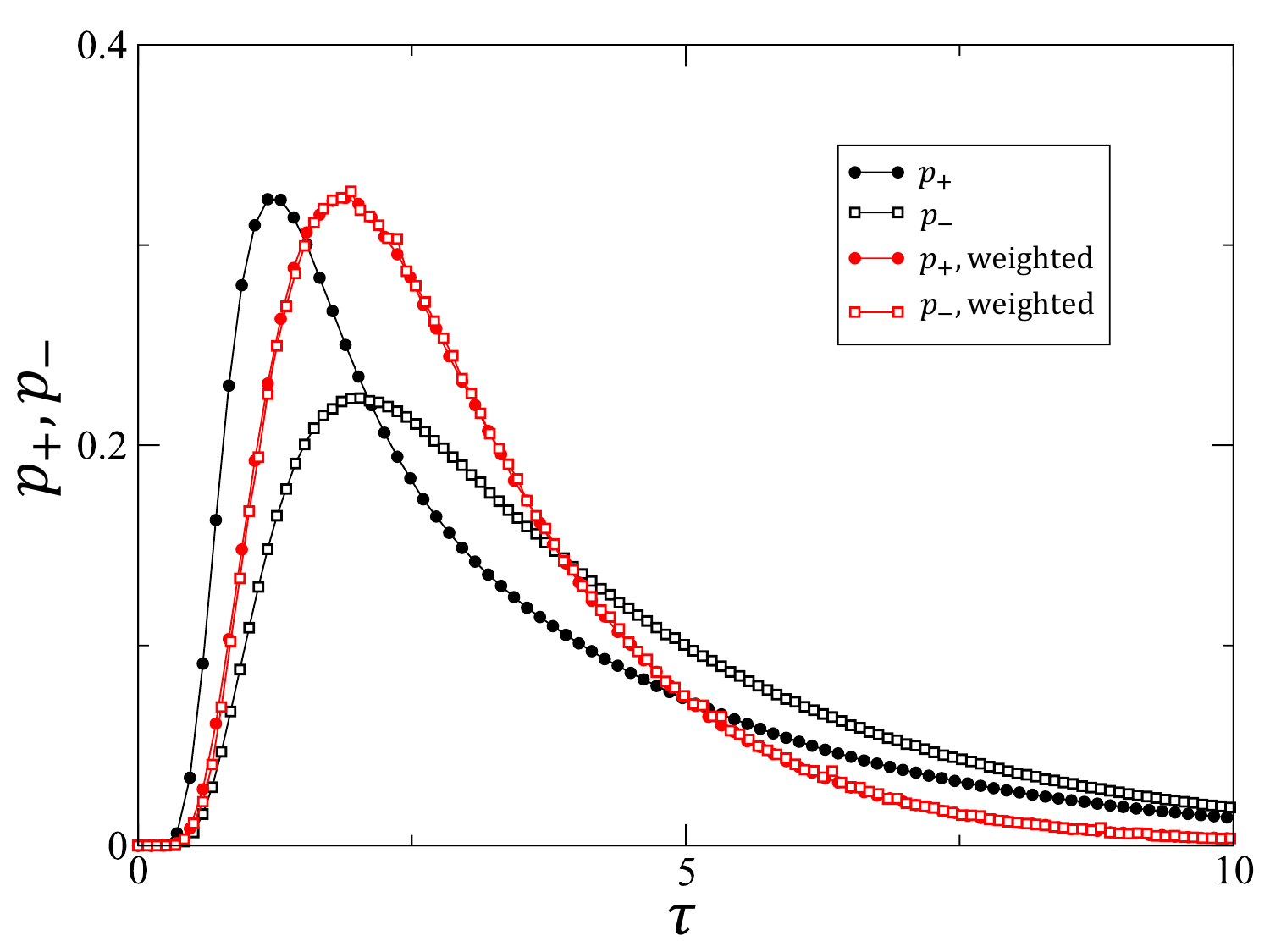}
\caption{Conditional first passage time probability
densities for downhill and uphill passages. Black filled
and open symbols denote the unweighted distributions
$p_+$ and $p_-$, whereas red filled and open symbols
denote the corresponding normalised half-weighted
distributions. Their agreement within numerical
resolution confirms the half-weighted equality for the
conditional first passage time distributions. The results
are shown for $U=0.26$ and $F=9.66$, at fixed
$\widetilde D=0.013$.}
\label{fig_WeightPDF}
\end{figure}

\begin{figure*}[htb]
\includegraphics[width=0.8\textwidth]{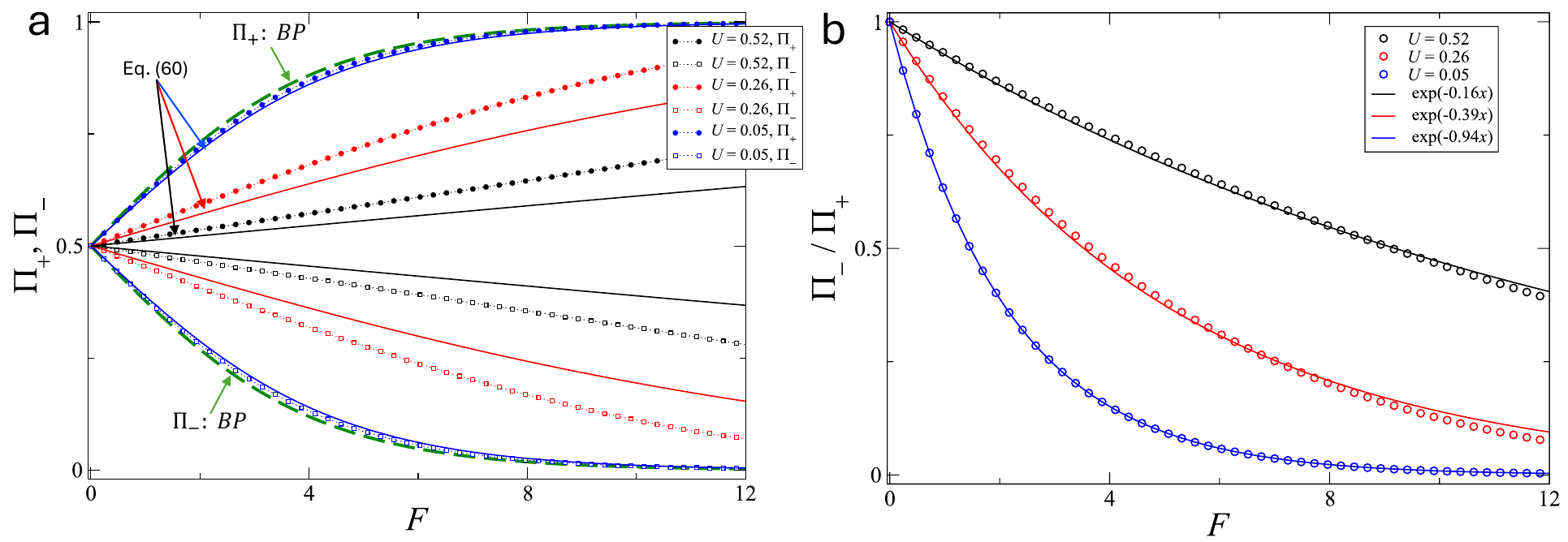}
\caption{(a) Splitting probabilities for downhill and uphill passages, \(\Pi_+\) and \(\Pi_-\), shown as functions of the dimensionless bias \(F\) for several values of the persistence parameter \(U\), at fixed \(\widetilde D =0.013\). The dashed green curves indicate the analytical passive Brownian reference behaviour, Eq.~\ref{eq:crossing_prob}, while the curves labelled Eq.~(\ref{eq:W_wp}) show the
effective drift-diffusion result whose expansion through $O(U^2)$ reproduces the 
contribution obtained from the weak-persistence expansion after retaining its leading term as $\varepsilon\to0$ at fixed $F$.
(b) Corresponding ratio \(\Pi_-/\Pi_+\) as a function of \(F\). The figure shows that for $F>0$ increasing persistence weakens the splitting asymmetry by shifting $\Pi_+$ and $\Pi_-$ towards 1/2, while driving systematic deviations from both the passive
result and the effective drift-diffusion result.}
\label{fig_crossing_prob}
\end{figure*}

\section{Numerical results: Directional asymmetry and path reweighting}
\label{sec:numerics}

Figures~\ref{fig_trajectories_2D} and
\ref{fig_trajectories} illustrate the different effects of
persistence and external bias on uphill and downhill
trajectories. Figure~\ref{fig_mean_times}(a) shows the
direction-conditioned mean first passage times
$\langle\tau_\sigma\rangle$ as functions of
$U=\ell_p/L$, for fixed $\widetilde D$ and several values
of $F$. The solid curves show the effective
drift-diffusion prediction from
Eq.~\eqref{eq:wp_conditional_mean}. Its $U\ll1$
expansion reproduces the common leading $O(U^2)$
contribution derived in Sec.~\ref{sec:5A}, but it does not
contain the boundary-sensitive remainders $R_\sigma$.
Within the weak-persistence expansion, the uphill--downhill
difference at $O(U^2)$ is therefore
$U^2(R_--R_+)$.

At $U=0$, the two direction-conditioned means coincide,
as they must for passive drift-diffusion starting from the centre of the interval. Their separation remains small for
$U\ll1$, in agreement with the common leading correction
generated by $D_{\rm eff}$. At larger $U$, both means
become non-monotonic functions of persistence, with a
stronger increase of $\langle\tau_-\rangle$. The ratio
$\langle\tau_-\rangle/\langle\tau_+\rangle$ in
Fig.~\ref{fig_mean_times}(b) therefore develops a maximum
that shifts to larger $U$ as $F$ increases. In this regime,
the orientation does not relax completely
during a typical passage, and a single effective
diffusivity is insufficient to describe both
direction-conditioned mean first passage times.

Figure~\ref{fig_mean_times_vsf} shows the corresponding
dependence on the external bias. For small $U$, both
direction-conditioned means decrease with increasing
$F$, close to the passive drift-diffusion result. In the
passive problem, successful uphill and downhill passages
have the same conditional mean duration, even though the
uphill splitting probability decreases strongly with
$F$.
At larger persistence, the two directions respond
differently. The downhill mean becomes only weakly
dependent on $F$, whereas the uphill mean increases over
most of the displayed range. As a result,
$\langle\tau_-\rangle/\langle\tau_+\rangle$ increases
approximately linearly with $F$, with a larger slope at
larger $U$.

The frozen-orientation approximation explains this
difference through the exit-dependent weights
$\Pi_\sigma(0,\phi_0)$. Increasing $F$ shifts the
projected drift
$v(\phi_0)=\widetilde D F+U\cos\phi_0$
towards positive values. Downhill passages remain
dominated by orientations with positive projected drift,
whereas the uphill weight $\Pi_-(0,\phi_0)$ increasingly
favours orientations whose active contribution opposes the
force. For these orientations, the net projected drift
becomes less negative and approaches $v=0$ as $F$
increases. Since the fixed-orientation conditional mean
first passage time is maximal at $v=0$ and decreases with
$|v|$, this increases $\langle\tau_-\rangle$ over the
force range shown, in agreement with
Eq.~\eqref{eq:frozen_orientation_averaged_mean}.
For $U=0.52$, the dashed curves in
Fig.~\ref{fig_mean_times_vsf}(a) show the
frozen-orientation estimates from
Eq.~\eqref{eq:frozen_orientation_averaged_mean}. Although
$U=0.52$ is not asymptotically large, the approximation
closely follows the numerical direction-conditioned mean
first passage times over the displayed range of $F$.

The conditional first passage time probability densities are shown in
Fig.~\ref{fig_PDFs_example}. At smaller $U$ in
Fig.~\ref{fig_PDFs_example}(a), $p_+(\tau)$ and
$p_-(\tau)$ remain close, with differences mainly in the
peak positions and long time tails. This agrees with the
weak persistence expansion, in which the leading
correction is common to the two directions.
As $U$ increases, 
Fig.~\ref{fig_PDFs_example}(b), the difference extends
over the full distributions. The downhill density
$p_+(\tau)$ peaks at a shorter time and is narrower,
whereas $p_-(\tau)$ has a broader long time tail. For
passive drift-diffusion, the corresponding conditional probability densities are identical,
despite their different splitting probabilities
\cite{dagbug:2009}. 

The difference between $p_+(\tau)$ and $p_-(\tau)$
can be understood in terms of the orientation histories
preferentially represented in the two conditioned
ensembles. Downhill passages are favoured by histories in
which the active propulsion supports the external drift
for a substantial part of the trajectory, whereas uphill
passages are favoured by histories in which it opposes the
force. Within this interpretation, the broader long-time
tail of $p_-(\tau)$ is associated with histories
for which the net projected drift remains small in
magnitude or changes sign during the passage, delaying
progress towards the uphill boundary.

These different orientation histories produce the visibly
different unweighted densities shown in
Fig.~\ref{fig_WeightPDF}. The same figure tests whether
the normalised half-weighting derived in
Sec.~\ref{sec:transition_paths} compensates for this
difference. After reweighting, the uphill and downhill
first passage time histograms agree within the numerical resolution
over both the peak and the long-time tail. 
Additional examples for weak and moderate persistence and for
different values of the bias are shown in
Appendix~\ref{app:additional_collapse},
Fig.~\ref{fig:appendix_weighted_collapse}.
The agreement of the half-weighted distributions persists
across these parameter regimes, even though the unweighted
uphill and downhill statistics differ. This equality does not rely on either the
weak-persistence expansion or the frozen-orientation
approximation; it follows directly from the
reflected-path relation derived in Sec.~\ref{sec:transition_paths}.

The splitting probabilities in
Fig.~\ref{fig_crossing_prob} provide a complementary
measure of the directional asymmetry. For a passive
particle released at the centre of the interval,
\begin{equation}
\Pi_+
=
\frac{1}{1+\exp(-F/2)},
\qquad
\Pi_-=1-\Pi_+ .
\label{eq:crossing_prob}
\end{equation}
The dashed green curves in
Fig.~\ref{fig_crossing_prob}(a) show this passive
reference.
The effective drift-diffusion result
Eq.~\eqref{eq:W_wp} is also shown in
Fig.~\ref{fig_crossing_prob}(a) by solid curves. The numerical data remain
close to the full effective expression at small $U$ while
deviating from it increasingly as $U$ grows. For every
$F>0$, Eq.~\eqref{eq:W_wp} shifts $\Pi_+$ below and
$\Pi_-$ above their passive values, moving both splitting
probabilities towards $1/2$.
The numerical results show
the same trend. Consequently, the ratio
$\Pi_-/\Pi_+$ in Fig.~\ref{fig_crossing_prob}(b) is larger
than its passive value and decreases more slowly with
$F$.
The frozen-orientation limit provides a complementary
interpretation of this behaviour. Uphill exits are then
associated predominantly with initial orientations for
which the projected drift
$v(\phi_0)=\widetilde D F+U\cos\phi_0$ is negative.
At fixed $U$, the angular range satisfying $v(\phi_0)<0$
decreases as $F$ increases, whereas at fixed $F$ it
increases with $U$. This accounts qualitatively for the
decrease of $\Pi_-/\Pi_+$ with increasing bias and for its
enhancement relative to the passive result as $U$ grows.
At finite $\widetilde D$, translational diffusion makes
this selection by the sign of the projected drift less
sharp, as described by
Eq.~\eqref{eq:frozen_orientation_splitting}.
Figure~\ref{fig_crossing_prob}(b) also compares the
splitting ratio with the exponential force dependence
expected if activity could be represented by a single
effective temperature, or equivalently by a scalar
effective diffusivity. The solid curves show exponential
fits to $\Pi_-/\Pi_+$. The data at the smallest persistence
are close to this form, whereas systematic deviations
appear as $U$ increases, particularly at larger $F$
\cite{preisler:2016,Cugliandolo:2019}. 

Finally, figures~\ref{fig_mean_times_vsf}(b) and
\ref{fig_crossing_prob}(b) emphasise that the two measures of
directional asymmetry respond oppositely to the external
bias. Over the displayed range,
$\langle\tau_-\rangle/\langle\tau_+\rangle$ increases
with $F$, whereas $\Pi_-/\Pi_+$ decreases. A stronger
bias therefore makes uphill exits less probable while
increasing the ratio of the uphill to downhill mean first
passage times.

\section{Experimental implications and possible protocols}
\label{sec:experimental}

\subsection{Reconstruction of uphill statistics from
downhill trajectories}

The reflected-path relation tested through the symmetric
half-weighting also gives a one-sided reconstruction of
the uphill first passage statistics from downhill
trajectories. Consider trajectories initiated at
$x=0$ and terminated when they first reach either of two
virtual boundaries at $x=\pm L/2$. To reproduce the
ensemble studied above, the initial orientations should
be isotropically distributed. For $F>0$, downhill exits
occur more frequently than uphill exits, so a fixed
observation time generally provides a larger downhill
sample.

Let $\omega_i^{(+)}$, $i=1,\ldots,N_+$, denote the
recorded downhill first passage trajectories, with first
passage times $\tau_i^{(+)}$. The uphill first passage time probability density
can be estimated by reweighting the recorded downhill trajectories 
\begin{equation}
\widehat p_-(\tau)
=
\frac{
\displaystyle
\sum_{i=1}^{N_+}
w_i\,
\delta_\Delta
\bigl(\tau-\tau_i^{(+)}\bigr)
}{
\displaystyle
\sum_{i=1}^{N_+}w_i
},
\qquad
w_i
=
\exp\!\left[-\Sigma[\omega_i^{(+)}]\right].
\label{eq:experimental_reweighting}
\end{equation}
The denominator implements the normalisation required to
convert averages over the conditional downhill ensemble
into averages over the conditional uphill ensemble. 
The function $\delta_\Delta$ is the top-hat representation
of the Dirac delta associated with a histogram bin of width
$\Delta$:
\begin{equation}
\delta_\Delta(z)
=
\begin{cases}
1/\Delta, & |z|<\Delta/2,\\[2mm]
0, & |z|\geq\Delta/2.
\end{cases}
\end{equation}

For downhill trajectories starting at $X=0$ and ending at
$X=1/2$, the force contribution to $\Sigma$ is the same
for every trajectory. The corresponding factor
$\exp(-F/2)$ therefore cancels between the numerator and
denominator of
Eq.~\eqref{eq:experimental_reweighting}. Reconstruction
of the normalised density $p_-(\tau)$ depends only on the
path-dependent part of $\Sigma$. 

Experimental reconstruction requires access to the
projected trajectory $X(t)$ and the orientation
$\phi(t)$ along each passage, together with the parameters
entering the path-dependent part of $\Sigma$. The
Stratonovich integral is then obtained from the measured
trajectory using the corresponding symmetric
discretisation. Thus, the reconstruction uses information
from the full transition path, rather than from the
first-passage time alone.
The statistical accuracy of the reconstruction also depends
on the distribution of the weights $w_i=e^{-\Sigma_i}$;
a broad weight distribution can make the estimator effectively
depend on only a small fraction of the recorded trajectories.
\subsection{Parameter inference and model validation}
Measurements of the uphill and downhill first passage
statistics provide a way to infer the parameters of an
experimental active-particle system. For a prescribed
interval length and an independently calibrated external
bias, the self-propulsion parameter $U$ and the
translational diffusivity $\widetilde D$ can be estimated
by fitting measured first passage observables to the
analytical or numerical predictions of the ABP model.
Useful observables include
\begin{equation}
R_\tau
=
\frac{\langle\tau_-\rangle}
{\langle\tau_+\rangle},
\qquad
R_\Pi
=
\frac{\Pi_-}{\Pi_+}.
\label{eq:experimental_observables}
\end{equation}
The two ratios contain complementary information:
$R_\tau$ measures the directional difference between the
conditional first passage times but is non-monotonic in $U$,
whereas $R_\Pi$ measures the asymmetry of the exit
probabilities. Measurements at several values of the
applied force or interval length can therefore be fitted
jointly to constrain $U$ and $\widetilde D$. If the external bias is not
independently known, $F$ can be included as an additional
fitted parameter.

The reconstruction of uphill statistics from downhill
trajectories provides a separate test of the inferred
parameters and of the ABP description. In a regime where some uphill passages can
still be measured directly, their first passage time
density can be compared with the density reconstructed
from downhill trajectories using
Eq.~\eqref{eq:experimental_reweighting}. Agreement would
support both the reflected-path relation and the
description of the experimental dynamics by the ABP
model. A systematic discrepancy, after accounting for
finite sampling, temporal resolution, and errors in the
measured orientation, would indicate limitations of the
ABP description or of the assumptions entering the
reconstruction.

\section{Conclusions}
\label{conclusion}

We studied direction-resolved first passage statistics of
an active Brownian particle driven by a constant external
force between two absorbing boundaries. For passive
drift-diffusion from the centre of the interval, the
conditional first passage time distributions for the two
exits are identical, whereas self-propulsion and
orientational persistence generate a directional
asymmetry between the corresponding uphill and downhill
distributions.

We derived an exact relation between the probability of a
trajectory $\omega$ and that of its spatially reflected
counterpart $\mathcal R\omega$. The logarithm of their
probability ratio defines the path-dependent asymmetry
functional $\Sigma[\omega]$, which determines the relative
probability of each reflected pair and the corresponding
reweighting factors. Reweighting the downhill and uphill
path probabilities by $\exp[-\Sigma/2]$ and
$\exp[+\Sigma/2]$, respectively, yields identical
half-weighted first passage time distributions for any
orientational persistence. The numerical results confirm
this equality over the range of first passage times
resolved in the simulations.
The same path-level relation also allows the uphill
conditional first passage time density to be reconstructed
from downhill trajectories by multiplying the probability
assigned to each downhill path by $\exp[-\Sigma]$.

The unweighted observables show that the directional
differences in the conditional first passage times and in
the exit probabilities have different dependences on
$U$. At fixed external bias, increasing $U$ moves the
splitting probabilities towards equal values.
The conditional first passage time statistics show a
different dependence, with a ratio of the conditional
mean first passage times that is non-monotonic in $U$ and
an uphill density that develops a broader long-time tail
at larger persistence. Over the force range studied,
$\langle\tau_-\rangle/\langle\tau_+\rangle$ increases
with $F$, whereas $\Pi_-/\Pi_+$ decreases. A stronger
bias therefore makes uphill exits less probable while
increasing the ratio of the uphill to downhill mean
first passage times.

The weak-persistence expansion and the frozen-orientation
approximation describe the origin of the directional
asymmetry in two different persistence regimes.
A weak-persistence expansion followed by the limit of rapid angular
relaxation shows that, at $O(U^2)$, the leading
contribution as $\varepsilon\to0$ at fixed $F$ comes from
the interior solution and is common to the two
directions. This contribution agrees with the effective
drift-diffusion result when the drift
$v=\varepsilon F$ is held fixed. The directional
difference at $O(U^2)$ arises from the absorbing-boundary
layers and is of relative order
$O(\sqrt{\varepsilon})$ or smaller compared with the
common interior contribution. The numerical solution
of the perturbative boundary-value hierarchy agrees with
the common leading asymptote.

In the frozen-orientation approximation, rotational
diffusion is negligible over the duration of a passage.
For a given initial orientation $\phi_0$, the
translational first passage problem then reduces to
drift-diffusion with a constant projected drift. When the
passage is conditioned on a particular exit, different
initial orientations contribute with weights determined by
the corresponding splitting probability
$\Pi_\sigma(0,\phi_0)$. Thus, although the conditional mean
at fixed $\phi_0$ is the same for the two boundaries, the
uphill and downhill means can differ after averaging over
the initial orientation. The approximation becomes less
accurate when the projected drift is close to zero,
because the passage time can then become comparable to the
time scale of angular relaxation.

The measured first passage data can also be used to
estimate model parameters and to test the reflected-path
relation experimentally. The ratios of the conditional
mean first passage times and of the splitting
probabilities, measured at several applied forces or
interval lengths, can be fitted jointly to constrain
parameters such as the self-propulsion parameter $U$ and
the dimensionless diffusivity $\widetilde D$.
If both the position and orientation histories are
resolved sufficiently well to evaluate $\Sigma$, the
measured downhill trajectories can also be used to
reconstruct the uphill first passage time density through
the reweighting factor $\exp[-\Sigma]$. When an independent
uphill distribution can be measured, its agreement with
the reconstructed density provides a direct test of the
ABP description and of the reflected-path relation.

The present construction relies on the reflection
$\mathcal R$, which maps the two directional first passage
ensembles onto one another while preserving the first
passage time. For systems with spatially varying forces,
absorbing boundaries at unequal distances from the
initial point, or interactions between active particles,
the corresponding path mapping and reweighting functional
must be identified for the modified dynamics and geometry.

\begin{figure*}[t]
\includegraphics[width=0.7\textwidth]{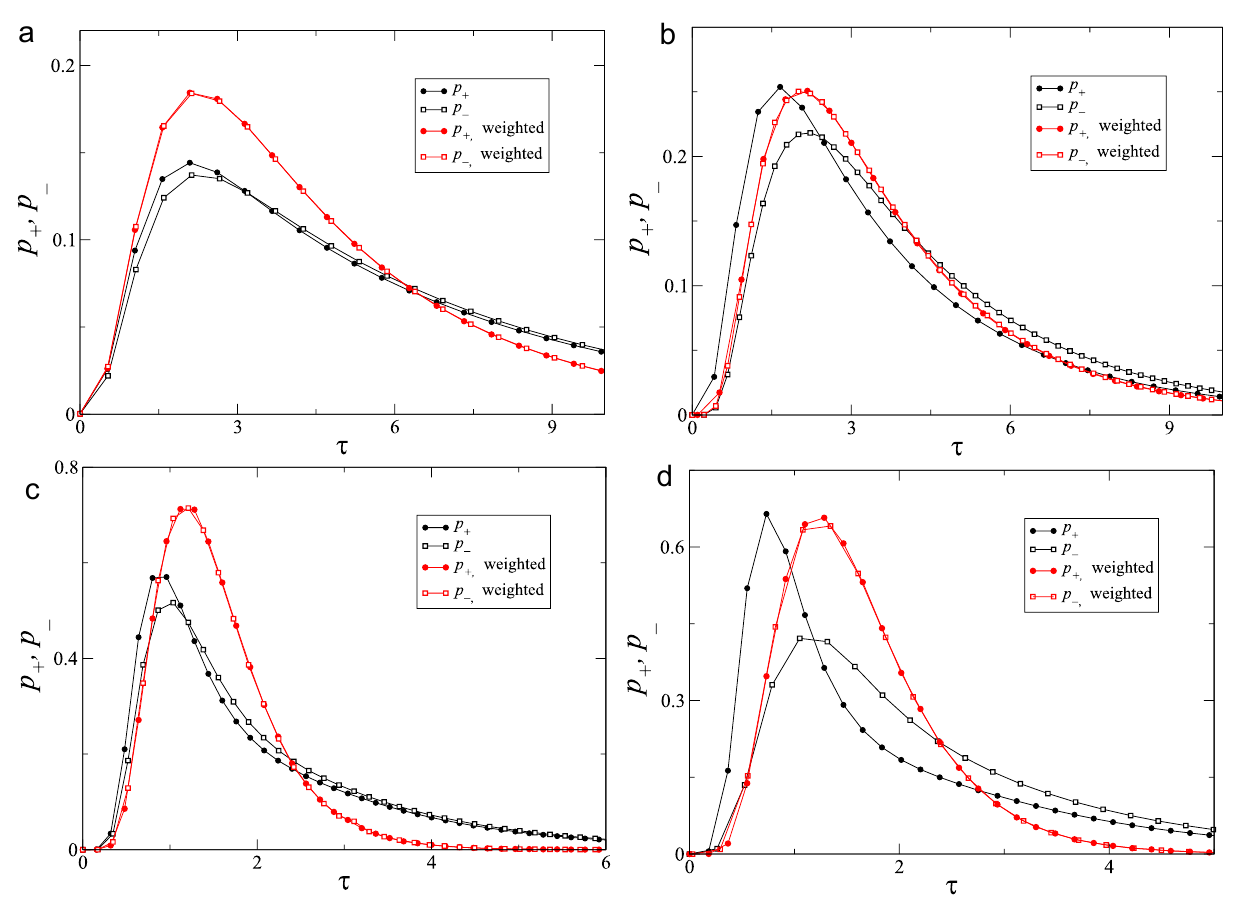}
\caption{Additional examples of the conditional first
passage time probability densities for representative
combinations of persistence and bias. Black filled and
open symbols denote the unweighted downhill and uphill
distributions, \(p_+(\tau)\) and \(p_-(\tau)\), while red
filled and open symbols denote the corresponding
normalised half-weighted distributions. Panels show:
(a) \(U=0.13\), \(F=2.41\); (b) \(U=0.13\),
\(F=9.66\); (c) \(U=0.52\), \(F=2.41\); and
(d) \(U=0.52\), \(F=9.66\). In all panels,
\(\widetilde D=0.013\). Although the unweighted uphill
and downhill conditional first passage time probability
densities differ substantially, the normalised
half-weighted distributions agree within numerical
resolution across all four representative regimes.}
\label{fig:appendix_weighted_collapse}
\end{figure*}


\appendix
\section{Additional numerical tests of the
half-weighted equality}
\label{app:additional_collapse}
Figure~\ref{fig:appendix_weighted_collapse} extends the
comparison in Fig.~\ref{fig_WeightPDF} to four additional
parameter sets: (a) $U=0.13$, $F=2.41$; (b) $U=0.13$,
$F=9.66$; (c) $U=0.52$, $F=2.41$; and (d) $U=0.52$,
$F=9.66$. These cases combine weak or moderate
persistence with low or high external bias.

For each parameter set, the unweighted conditional first
passage time densities $p_+(\tau)$ and $p_-(\tau)$ are
visibly different. After applying the normalised
half-weighting defined in
Sec.~\ref{sec:transition_paths}, the corresponding
weighted densities agree within the numerical resolution.
The agreement is therefore observed for all four
combinations of persistence and bias shown here, in
addition to the example presented in the main text.

\section{Stochastic simulations}
\label{app:numerics}

The dimensionless projected dynamics for $(X,\phi)$ were
integrated using the Euler--Maruyama scheme. With
$\tau_n=n\Delta\tau$, the updates are
\begin{align}
X_{n+1}
&=
X_n
+
\left(
U\cos\phi_n+\widetilde D F
\right)\Delta\tau
+
\sqrt{2\widetilde D\,\Delta\tau}\,
\eta_n^{(X)},
\\
\phi_{n+1}
&=
\phi_n
+
\sqrt{2\Delta\tau}\,
\eta_n^{(\phi)},
\end{align}
where $\eta_n^{(X)}$ and $\eta_n^{(\phi)}$ are independent
standard normal random variables. Unless stated otherwise,
we used $\Delta\tau=5\times10^{-4}$. Time-step sensitivity
was checked by repeating representative calculations with
$\Delta\tau=10^{-4}$, $10^{-3}$, and $5\times10^{-3}$;
the reported trends were unchanged.

Each first passage run started from $X_0=0$, with
$\phi_0$ sampled independently from the uniform
distribution on $[0,2\pi)$. The run was terminated at the
first discrete time step for which $X_n\geq1/2$ or
$X_n\leq-1/2$. The corresponding event was classified as
downhill or uphill, respectively. No interpolation of the
crossing time within the final time step was performed.

All stochastic-simulation results in the main text and in
Appendix~\ref{app:additional_collapse} were obtained at
fixed $\widetilde D=0.013$. We considered five values of
$F$ in the range $2\lesssim F\lesssim12$ and, for each
$F$, seven values of $U$ in the range
$0.05\leq U\leq1.05$. For each parameter set, the statistics were accumulated
from approximately $10^{10}$ downhill and $10^9$ uphill
first passage events.

The path-dependent asymmetry functional was evaluated
along each trajectory using the same discrete increments
as in the dynamical update. Numerically, both the force
and active contributions were retained in the stochastic
integral,
\begin{equation}
\Sigma[\omega]
\simeq
\sum_{n=1}^{N}
\left[
F
+
\frac{U}{\widetilde D}
\frac{\cos\phi_n+\cos\phi_{n-1}}{2}
\right]
\Delta X_n,
\label{eq:discrete_stratonovich}
\end{equation}
where $N$ is the first time step at which an absorbing
boundary is crossed, and $\Delta X_n=X_n-X_{n-1}$. The final increment
$\Delta X_N$ was retained in the sum, including the
overshoot beyond the absorbing boundary.

For the half-weighted distributions in
Figs.~\ref{fig_WeightPDF} and
\ref{fig:appendix_weighted_collapse}, each trajectory
$\omega$ was assigned the weight
\begin{equation}
\exp\!\left[-\frac12\Sigma[\omega]\right].
\end{equation}
The uphill and downhill weighted histograms were
normalised separately. Since $\Sigma[\mathcal R\omega]=-\Sigma[\omega]$, a
downhill path $\omega\in\Omega_+(\tau)$ and its reflected
uphill partner $\mathcal R\omega\in\Omega_-(\tau)$ receive
the weights
$\exp[-\Sigma[\omega]/2]$ and
$\exp[-\Sigma[\mathcal R\omega]/2]
=\exp[+\Sigma[\omega]/2]$, respectively.
\section{Weak-persistence expansion of the direction-conditioned mean first passage times}
\label{app:weak_persistence}

This appendix derives the leading weak-persistence
correction to $\langle\tau_\sigma\rangle$. Here and below, 
the expansion in $U$ is performed  before
taking the rapid angular relaxation limit of its
coefficients. We use the
splitting probability $\Pi_\sigma(X_0,\phi_0)$ and the
unnormalised first directional moment
$M_\sigma(X_0,\phi_0)$ defined in the main text. Their
orientational averages determine the direction-conditioned
mean through Eq.~\eqref{eq:conditional_mean_from_moments},
$\langle\tau_\sigma\rangle=M_\sigma/\Pi_\sigma$.
The functions $\Pi_\sigma(X_0,\phi_0)$ and
$M_\sigma(X_0,\phi_0)$ satisfy the backward equations
\eqref{eq:splitting_backward} and
\eqref{eq:directional_moment_backward}, together with the
boundary conditions \eqref{eq:Pi_sigma_bc} and
\eqref{eq:directional_moment_bc}, respectively.

We expand $\Pi_\sigma$ and $M_\sigma$ in powers of $U$
and retain terms through $O(U^2)$. The $O(U)$ terms are
first angular harmonics and vanish upon averaging over the
initial orientation, whereas the leading correction to the
orientationally averaged quantities appears at $O(U^2)$.
At $O(U)$, the first angular harmonics are
$U\pi_{1,\sigma}\cos\phi_0$ and
$Um_{1,\sigma}\cos\phi_0$, respectively. At $O(U^2)$,
the self-propulsion term
$U\cos\phi_0\,\partial_{X_0}$ acts on these first
harmonics. Since $\cos^2\phi_0 = (1+\cos(2\phi_0))/2$,
the derivatives
$\partial_{X_0}\pi_{1,\sigma}$ and
$\partial_{X_0}m_{1,\sigma}$ enter both the equations for
the angle-independent coefficients
$\pi_{20,\sigma}$ and $m_{20,\sigma}$ and the equations
for the coefficients multiplying $\cos(2\phi_0)$.

Only the angle-independent $O(U^2)$ coefficients
$\pi_{20,\sigma}$ and $m_{20,\sigma}$ survive the
orientational averaging of $\Pi_\sigma$ and $M_\sigma$.
Their equations, however, contain
$\partial_{X_0}\pi_{1,\sigma}$ and
$\partial_{X_0}m_{1,\sigma}$ as source terms.
Determining the orientationally averaged quantities at
$O(U^2)$ therefore requires the $O(U)$ first harmonics.

In the rapid angular relaxation limit, the first-harmonic
amplitudes $\pi_{1,\sigma}$ and $m_{1,\sigma}$ consist of
bulk parts, valid away from the absorbing endpoints, and
boundary-layer corrections required to satisfy the 
boundary conditions. Substituting the bulk parts of these
amplitudes into the angle-independent $O(U^2)$ equations
gives the leading contributions to
$\pi_{20,\sigma}$ and $m_{20,\sigma}$ in this limit.
The boundary-layer parts
of the first harmonics generate additional contributions to
$\pi_{20,\sigma}$ and $m_{20,\sigma}$. These contributions are
subleading in the rapid angular relaxation expansion and
produce the directional correction discussed below.

The terms generated by the bulk parts of
$\pi_{1,\sigma}$ and $m_{1,\sigma}$ can be expressed as
derivatives of the corresponding passive solutions with
respect to $\widetilde D$ at fixed drift
$v=\widetilde D F$. After forming
$M_\sigma/\Pi_\sigma$ and evaluating it at $X_0=0$, this
derivative representation provides the leading
rapid angular relaxation contribution to the $O(U^2)$
correction to $\langle\tau_\sigma\rangle$. Since the
passive direction-conditioned mean first passage times
from the centre of the interval are identical for the two
exits, this leading contribution is the same for
$\sigma=+1$ and $\sigma=-1$. The difference between the
two directions at $O(U^2)$ arises from the boundary-layer
terms in $\pi_{1,\sigma}$ and $m_{1,\sigma}$.

\subsection{Expansion in powers of $U$}
\label{appC:exact_expansion}

We denote the spatial part of the backward operator
\eqref{eq:backward_operator} by
\begin{equation}
\mathcal A
=
\widetilde D\,\partial_{X_0}^{2}
+
\widetilde D F\,\partial_{X_0},
\label{eq:appC_A}
\end{equation}
and use a prime to denote differentiation with respect to
the initial position $X_0$.

The backward equations and their boundary conditions are
invariant under the reflection
$\phi_0\mapsto-\phi_0$ (modulo $2\pi$). By uniqueness of
the corresponding periodic boundary-value problems,
$\Pi_\sigma(X_0,\phi_0)$ and $M_\sigma(X_0,\phi_0)$ are
even functions of $\phi_0$. Their angular Fourier
expansions therefore contain only the constant mode and
cosine harmonics. The self-propulsion term in the backward
operator is proportional to
$U\cos\phi_0\,\partial_{X_0}$. Because the passive
solutions are independent of $\phi_0$, this term
generates a $\cos\phi_0$ component at $O(U)$. Acting once
more on this first harmonic at $O(U^2)$ produces an
angle-independent component and a second harmonic
$\cos(2\phi_0)$.
We therefore expand the splitting probability as
\begin{equation}
\begin{split}
\Pi_\sigma(X_0,\phi_0)
={}&
\pi_{0,\sigma}(X_0)
+
U\pi_{1,\sigma}(X_0)\cos\phi_0
\\
&+
U^2
\left[
\pi_{20,\sigma}(X_0)
+
\pi_{22,\sigma}(X_0)\cos(2\phi_0)
\right]
\\
&+
O(U^3).
\end{split}
\label{eq:appC_Pi_expansion}
\end{equation}
Similarly, the unnormalised first directional moment is
expanded as
\begin{equation}
\begin{split}
M_\sigma(X_0,\phi_0)
={}&
m_{0,\sigma}(X_0)
+
Um_{1,\sigma}(X_0)\cos\phi_0
\\
&+
U^2
\left[
m_{20,\sigma}(X_0)
+
m_{22,\sigma}(X_0)\cos(2\phi_0)
\right]
\\
&+
O(U^3).
\end{split}
\label{eq:appC_M_expansion}
\end{equation}
Only the coefficients $\pi_{0,\sigma}$,
$\pi_{20,\sigma}$, $m_{0,\sigma}$, and
$m_{20,\sigma}$ survive the orientational average over
$\phi_0$. The first-harmonic amplitudes
$\pi_{1,\sigma}$ and $m_{1,\sigma}$ are nevertheless
required because their spatial derivatives appear as
source terms in the equations for $\pi_{20,\sigma}$ and
$m_{20,\sigma}$. The second-harmonic coefficients
$\pi_{22,\sigma}$ and $m_{22,\sigma}$ integrate to zero
and are therefore not needed for
$\langle\tau_\sigma\rangle$ through $O(U^2)$.

The passive splitting probabilities satisfy the
 boundary conditions
\begin{equation}
\begin{aligned}
\pi_{0,+}\!\left(-\frac12\right)&=0,
&
\pi_{0,+}\!\left(\frac12\right)&=1,
\\
\pi_{0,-}\!\left(-\frac12\right)&=1,
&
\pi_{0,-}\!\left(\frac12\right)&=0.
\end{aligned}
\label{eq:appC_pi0_BC}
\end{equation}
Because these boundary values are independent of $U$,
all higher-order coefficients in
Eq.~\eqref{eq:appC_Pi_expansion} vanish at both absorbing
boundaries. Since $M_\sigma(X_0,\phi_0)$ itself vanishes at both
boundaries for all $U$, every coefficient in
Eq.~\eqref{eq:appC_M_expansion} satisfies homogeneous
boundary conditions.

Substitution of Eq.~\eqref{eq:appC_Pi_expansion} into the
backward equation \eqref{eq:splitting_backward} for $\Pi_\sigma(X_0,\phi_0)$, followed by projection
onto the relevant angular harmonics, gives
\begin{subequations}
\label{eq:appC_pi_hierarchy}
\begin{align}
\mathcal A\pi_{0,\sigma}
&=0,
\label{eq:appC_pi0}
\\
(1-\mathcal A)\pi_{1,\sigma}
&=
\pi_{0,\sigma}',
\label{eq:appC_pi1}
\\
\mathcal A\pi_{20,\sigma}
&=
-\frac12\pi_{1,\sigma}'.
\label{eq:appC_pi20}
\end{align}
\end{subequations}
The corresponding equations obtained from the backward
equation \eqref{eq:directional_moment_backward} for $M_\sigma(X_0,\phi_0)$ are
\begin{subequations}
\label{eq:appC_m_hierarchy}
\begin{align}
\mathcal A m_{0,\sigma}
&=
-\pi_{0,\sigma},
\label{eq:appC_m0}
\\
(1-\mathcal A)m_{1,\sigma}
&=
m_{0,\sigma}'
+
\pi_{1,\sigma},
\label{eq:appC_m1}
\\
\mathcal A m_{20,\sigma}
&=
-\pi_{20,\sigma}
-\frac12m_{1,\sigma}'.
\label{eq:appC_m20}
\end{align}
\end{subequations}
Equations~\eqref{eq:appC_pi_hierarchy} and
\eqref{eq:appC_m_hierarchy} follow directly by matching
powers of $U$ and the corresponding angular harmonics; no
approximation in $\varepsilon$ has been introduced at this
stage. They therefore give the complete coefficient
hierarchy required to determine
$\langle\tau_\sigma\rangle$ through $O(U^2)$ at fixed
$\widetilde D$ and $F$. The rapid angular relaxation
limit is introduced only in the next subsection.

\subsection{Rapid angular relaxation asymptotics}
\label{appC:rapid_relaxation}

We now determine the leading behaviour of the coefficients
in Eqs.~\eqref{eq:appC_pi_hierarchy} and
\eqref{eq:appC_m_hierarchy} when angular relaxation is fast
compared with the projected spatial dynamics. We set
\begin{equation}
\varepsilon\equiv\widetilde D,
\qquad
\mathcal A
=
\varepsilon\mathcal A^{(0)},
\qquad
\mathcal A^{(0)}
=
\partial_{X_0}^{2}
+
F\partial_{X_0},
\label{eq:appC_A0}
\end{equation}
and consider the limit $\varepsilon\to0$ at fixed $F$.

The amplitudes $\pi_{1,\sigma}$ and $m_{1,\sigma}$
multiply the first angular harmonic $\cos\phi_0$, for
which
$\partial_{\phi_0}^{2}\cos\phi_0=-\cos\phi_0$.
Its relaxation rate in the dimensionless variables is
therefore unity, whereas the terms in $\mathcal A$ arising from
translational diffusion and force-induced drift are
proportional to $\varepsilon$ at fixed $F$.

Equations~\eqref{eq:appC_pi20} and
\eqref{eq:appC_m20} show that
$\partial_{X_0}\pi_{1,\sigma}$ and
$\partial_{X_0}m_{1,\sigma}$ enter as source terms in the
equations for the angle-independent $O(U^2)$ coefficients
$\pi_{20,\sigma}$ and $m_{20,\sigma}$. We therefore first
determine the behaviour of the two first-harmonic
amplitudes for $\varepsilon\to0$. Away from the absorbing
endpoints they have regular interior solutions, while
narrow boundary-layer corrections are required to satisfy
their homogeneous boundary conditions. The contributions
of these two parts to $\pi_{20,\sigma}$ and
$m_{20,\sigma}$ are compared below.

\subsubsection{First harmonic amplitude of the splitting probability}

\noindent Using \eqref{eq:appC_A0}, Eq.~\eqref{eq:appC_pi1} may be written as
\begin{equation}
\left(
1-\varepsilon\mathcal A^{(0)}
\right)
\pi_{1,\sigma}
=
\pi_{0,\sigma}'.
\label{eq:appC_pi1_eps}
\end{equation}
The passive splitting probability satisfies
\begin{equation*}
\mathcal A^{(0)}\pi_{0,\sigma}=0.
\end{equation*}
Since $\mathcal A^{(0)}$ has constant coefficients, it
commutes with differentiation with respect to $X_0$.
Therefore,
\begin{equation}
\mathcal A^{(0)}\pi_{0,\sigma}'
=
\left(
\mathcal A^{(0)}\pi_{0,\sigma}
\right)'
=
0.
\label{eq:appC_pi0_prime}
\end{equation}
It follows that $\pi_{0,\sigma}'$ satisfies the
differential equation in
Eq.~\eqref{eq:appC_pi1_eps}. It does not, in general,
satisfy the homogeneous boundary conditions required for
$\pi_{1,\sigma}$. We therefore introduce a boundary correction 
\begin{equation}
q_{\pi,\sigma}
\equiv
\pi_{1,\sigma}
-
\pi_{0,\sigma}',
\label{eq:appC_qpi_def}
\end{equation}
and exploiting Eq.~\eqref{eq:appC_pi1_eps} we obtain
\begin{equation}
\left(
1-\varepsilon\mathcal A^{(0)}
\right)
q_{\pi,\sigma}
=
0,
\label{eq:appC_qpi_eq}
\end{equation}
while the homogeneous boundary conditions on
$\pi_{1,\sigma}$ imply
\begin{equation}
q_{\pi,\sigma}\!\left(\pm\frac12\right)
=
-\pi_{0,\sigma}'\!\left(\pm\frac12\right).
\label{eq:appC_qpi_BC}
\end{equation}
Explicitly, Eq.~\eqref{eq:appC_qpi_eq} reads
\begin{equation}
\varepsilon q_{\pi,\sigma}''
+
\varepsilon F q_{\pi,\sigma}'
-
q_{\pi,\sigma}
=
0.
\label{eq:appC_qpi_explicit}
\end{equation}
The corresponding characteristic equation is
\begin{equation}
\varepsilon r^2+\varepsilon F r-1=0,
\end{equation}
with roots
\begin{equation}
r_{\pm}
=
-\frac{F}{2}
\pm
\frac{1}{\sqrt{\varepsilon}}
\sqrt{1+\frac{\varepsilon F^2}{4}}.
\label{eq:appC_qpi_roots}
\end{equation}
For $\varepsilon\to0$,
\begin{equation}
r_{\pm}
=
-\frac{F}{2}
\pm
\frac{1}{\sqrt{\varepsilon}}
+
O(\sqrt{\varepsilon}).
\end{equation}
The dominant contribution to the characteristic roots is
therefore of order $\varepsilon^{-1/2}$, which sets the
inverse spatial length scale of the boundary correction.
If $d$ denotes the distance measured into the interval
from the corresponding absorbing boundary, the decaying
solution behaves to leading order as
\begin{equation}
q_{\pi,\sigma}
\propto
\exp\!\left(-\frac{d}{\sqrt{\varepsilon}}\right).
\end{equation}
The $F$-dependent $O(1)$ contribution to the characteristic
roots changes the detailed boundary-layer profile but not its
leading width,
\begin{equation}
\delta=O(\sqrt{\varepsilon}).
\label{eq:appC_qpi_width}
\end{equation}

The boundary values of $q_{\pi,\sigma}$ in
Eq.~\eqref{eq:appC_qpi_BC} remain $O(1)$.
Consequently, at any fixed point in the interior, the
contributions propagated from the two endpoints are
exponentially small as $\varepsilon\to0$.
At the centre $X_0=0$, the distance to either boundary
is $1/2$, and therefore
\begin{equation}
q_{\pi,\sigma}(0)
=
O\!\left(
e^{-1/(2\sqrt{\varepsilon})}
\right).
\label{eq:appC_qpi_center}
\end{equation}
 Hence
\begin{equation}
\pi_{1,\sigma}(0)
=
\pi_{0,\sigma}'(0)
+
O\!\left(
e^{-1/(2\sqrt{\varepsilon})}
\right).
\label{eq:appC_pi1_center}
\end{equation}
More generally, whenever the distance $d$ from the
nearest endpoint satisfies
$d/\sqrt{\varepsilon}\to\infty$, the first-harmonic
amplitude $\pi_{1,\sigma}$ approaches
$\pi_{0,\sigma}'$ exponentially rapidly. The correction
$q_{\pi,\sigma}$ becomes $O(1)$ within the
$O(\sqrt{\varepsilon})$ boundary layers.

\subsubsection{First harmonic amplitude of the directional moment}

\noindent
Using Eq.~\eqref{eq:appC_A0}, Eq.~\eqref{eq:appC_m1}
becomes
\begin{equation}
\left(
1-\varepsilon\mathcal A^{(0)}
\right)
m_{1,\sigma}
=
m_{0,\sigma}'
+
\pi_{1,\sigma}.
\label{eq:appC_m1_eps}
\end{equation}
The passive directional moment satisfies
\begin{equation}
\varepsilon\mathcal A^{(0)}m_{0,\sigma}
=
-\pi_{0,\sigma}.
\label{eq:appC_m0_eps}
\end{equation}
Differentiation of the lhs with respect to $X_0$ gives
$\varepsilon\mathcal A^{(0)}m_{0,\sigma}'
=-\pi_{0,\sigma}'$, and therefore
\begin{equation}
\left(
1-\varepsilon\mathcal A^{(0)}
\right)
m_{0,\sigma}'
=
m_{0,\sigma}'
+
\pi_{0,\sigma}'.
\label{eq:appC_m0_prime_eq}
\end{equation}
Using
$\pi_{1,\sigma}=\pi_{0,\sigma}'+q_{\pi,\sigma}$,
the right-hand sides of
Eqs.~\eqref{eq:appC_m1_eps} and
\eqref{eq:appC_m0_prime_eq} differ by
$q_{\pi,\sigma}$.  Moreover, $m_{0,\sigma}'$ does not satisfy the homogeneous boundary conditions
required for $m_{1,\sigma}$. We thus introduce another boundary-correcting term
\begin{equation}
q_{m,\sigma}
\equiv
m_{1,\sigma}
-
m_{0,\sigma}'.
\label{eq:appC_qm_def}
\end{equation}
Subtracting Eq.~\eqref{eq:appC_m0_prime_eq} from
Eq.~\eqref{eq:appC_m1_eps} then provides
\begin{equation}
\left(
1-\varepsilon\mathcal A^{(0)}
\right)
q_{m,\sigma}
=
q_{\pi,\sigma}.
\label{eq:appC_qm_eq}
\end{equation}
Since $m_{1,\sigma}$ satisfies homogeneous boundary
conditions,
\begin{equation}
q_{m,\sigma}\!\left(\pm\frac12\right)
=
-
m_{0,\sigma}'\!\left(\pm\frac12\right).
\label{eq:appC_qm_BC}
\end{equation}

To determine the boundary-layer behaviour of
$q_{m,\sigma}$, we note from
Eq.~\eqref{eq:appC_m0_eps} that
$m_{0,\sigma}$ and $m_{0,\sigma}'$ are
$O(\varepsilon^{-1})$.
The boundary values of $q_{m,\sigma}$ in Eq.~\eqref{eq:appC_qm_BC} are
therefore also $O(\varepsilon^{-1})$.
The operator acting on $q_{m,\sigma}$ in
Eq.~\eqref{eq:appC_qm_eq} is the same operator that
governs $q_{\pi,\sigma}$ in
Eq.~\eqref{eq:appC_qpi_eq}. Its homogeneous solutions
therefore have the characteristic exponents given by
Eq.~\eqref{eq:appC_qpi_roots} and decay away from the
absorbing endpoints over a distance
$O(\sqrt{\varepsilon})$. Thus, a boundary value of
magnitude $O(\varepsilon^{-1})$ produces, at a distance
$d$ from that boundary, a contribution of order
\begin{equation*}
O\!\left[
\varepsilon^{-1}
\exp\!\left(
-\frac{d}{\sqrt{\varepsilon}}
\right)
\right].
\end{equation*}
The factor $\varepsilon^{-1}$ is inherited from the
boundary value itself, while the exponential factor
describes its decay from the boundary into the interval.
The source $q_{\pi,\sigma}$ on the rhs of
Eq.~\eqref{eq:appC_qm_eq} is confined to the same
$O(\sqrt{\varepsilon})$ boundary layers and is $O(1)$ there.
It
therefore modifies the detailed form of $q_{m,\sigma}$
within these layers without changing its leading
 scaling behaviour. Therefore, at the interval centre, where $d=1/2$ for both boundaries, we have
\begin{equation}
q_{m,\sigma}(0)
=
O\!\left[
\varepsilon^{-1}
\exp\!\left(
-\frac{1}{2\sqrt{\varepsilon}}
\right)
\right],
\label{eq:appC_qm_center}
\end{equation}
and using Eq.~\eqref{eq:appC_qm_def}, we obtain
\begin{equation}
m_{1,\sigma}(0)
=
m_{0,\sigma}'(0)
+
O\!\left[
\varepsilon^{-1}
\exp\!\left(
-\frac{1}{2\sqrt{\varepsilon}}
\right)
\right].
\label{eq:appC_m1_center}
\end{equation}
Thus $m_{1,\sigma}(0)$ approaches
$m_{0,\sigma}'(0)$ exponentially rapidly as
$\varepsilon\to0$.

\subsubsection{Angle independent $O(U^2)$ coefficients}

\noindent Substituting 
\begin{equation*}
\pi_{1,\sigma}'
=
\pi_{0,\sigma}''
+
q_{\pi,\sigma}',
\qquad
m_{1,\sigma}'
=
m_{0,\sigma}''
+
q_{m,\sigma}',
\end{equation*}
into Eqs.~\eqref{eq:appC_pi20} and
\eqref{eq:appC_m20} we obtain
\begin{equation}
\mathcal A\pi_{20,\sigma}
=
-\frac12 \left ( \pi_{0,\sigma}'' + q_{\pi,\sigma}' \right),
\label{eq:appC_pi20_split}
\end{equation}
and
\begin{equation}
\mathcal A m_{20,\sigma}
=
-\pi_{20,\sigma}
-\frac12 \left ( m_{0,\sigma}'' + q_{m,\sigma}' \right ).
\label{eq:appC_m20_split}
\end{equation}

Equations~\eqref{eq:appC_pi20_split} and
\eqref{eq:appC_m20_split} separate the terms involving
the boundary corrections $q_{\pi,\sigma}$ and
$q_{m,\sigma}$ from the spatially extended sources.
In Eq.~\eqref{eq:appC_m20_split},
$\pi_{20,\sigma}$ itself contains both contributions and
will be decomposed below.
The preceding two subsections showed that
$q_{\pi,\sigma}=O(1)$ and
$q_{m,\sigma}=O(\varepsilon^{-1})$ within layers of width
$O(\sqrt{\varepsilon})$ adjacent to the absorbing
boundaries, and that both corrections decay exponentially
into the interior.

We now estimate their contributions to
$\pi_{20,\sigma}$ and $m_{20,\sigma}$. Let
$G^{(0)}(X_0,X')$ denote the Dirichlet Green's function of
$\mathcal A^{(0)}$. Since
$\mathcal A=\varepsilon\mathcal A^{(0)}$, the solution of
$\mathcal A u=f$ with homogeneous boundary conditions can
be written as
\begin{equation*}
u(X_0)
=
\frac{1}{\varepsilon}
\int_{-1/2}^{1/2}
G^{(0)}(X_0,X')f(X')\,dX'.
\end{equation*}
For $X_0=0$, the Green's function vanishes when $X'$ reaches
either absorbing boundary, while
$\partial_{X'}G^{(0)}(0,X')$ remains bounded in their
neighbourhoods.

We denote by $\pi_{20,\sigma}^{\rm BL}$ (BL from "Boundary Layer") the contribution
generated by $q_{\pi,\sigma}'$. From
Eq.~\eqref{eq:appC_pi20_split},
\begin{equation*}
\pi_{20,\sigma}^{\rm BL}(0)
=
-\frac{1}{2\varepsilon}
\int_{-1/2}^{1/2}
G^{(0)}(0,X')q_{\pi,\sigma}'(X')\,dX'.
\end{equation*}
Integration by parts gives
\begin{equation*}
\pi_{20,\sigma}^{\rm BL}(0)
=
\frac{1}{2\varepsilon}
\int_{-1/2}^{1/2}
\partial_{X'}G^{(0)}(0,X')
q_{\pi,\sigma}(X')\,dX',
\end{equation*}
since $G^{(0)}(0,X')$ vanishes at both boundaries.
The correction $q_{\pi,\sigma}$ is appreciable only
within boundary layers of width
$O(\sqrt{\varepsilon})$ and is $O(1)$ there. Since
$\partial_{X'}G^{(0)}(0,X')$ remains finite near the
boundaries, the integral is therefore $O(\sqrt{\varepsilon})$. Together with the
prefactor $1/\varepsilon$, this gives
\begin{equation}
\pi_{20,\sigma}^{\rm BL}(0)
=
O(\varepsilon^{-1/2}).
\label{eq:appC_pi20_BL_scaling}
\end{equation}

We now consider the boundary-layer contribution $m_{20,\sigma}^{\rm BL}$, which obeys 
\begin{equation}
\mathcal A m_{20,\sigma}^{\rm BL}
=
-\pi_{20,\sigma}^{\rm BL}
-\frac12 q_{m,\sigma}',
\label{eq:appC_m20_BL}
\end{equation}
with homogeneous boundary conditions.
The part resulting from $q_{m,\sigma}'$ can be estimated using
the same Green-function argument as for
$\pi_{20,\sigma}^{\rm BL}$, and at the
interval centre it reads
\begin{equation*}
\frac{1}{2\varepsilon}
\int_{-1/2}^{1/2}
\partial_{X'}G^{(0)}(0,X')
q_{m,\sigma}(X')\,dX'.
\end{equation*}
As $q_{m,\sigma}=O(\varepsilon^{-1})$ within
$O(\sqrt{\varepsilon})$ boundary layers and decays
exponentially into the interior, the expression above is
$O(\varepsilon^{-3/2})$.
The term $-\pi_{20,\sigma}^{\rm BL}$ in
Eq.~\eqref{eq:appC_m20_BL} gives a contribution of the
same order. Since  
$\pi_{20,\sigma}^{\rm BL}(X_0)=O(\varepsilon^{-1/2})$
throughout the interval, 
\begin{equation*}
-\frac{1}{\varepsilon}
\int_{-1/2}^{1/2}
G^{(0)}(0,X')\,
\pi_{20,\sigma}^{\rm BL}(X')\,dX'
=
O(\varepsilon^{-3/2}),
\end{equation*}
with $G^{(0)}(0,X')$ finite on the integration interval.
Hence
\begin{equation}
m_{20,\sigma}^{\rm BL}(0)
=
O(\varepsilon^{-3/2}).
\label{eq:appC_m20_BL_scaling}
\end{equation}

We now compare these boundary layer generated
contributions to $\pi_{20,\sigma}$ and $m_{20,\sigma}$ with those produced by the spatially
extended source terms. We denote by
$\pi_{20,\sigma}^{\mathrm{blk}}$ the contribution
generated by $-\pi_{0,\sigma}''/2$ through
\begin{equation}
\mathcal A\pi_{20,\sigma}^{\mathrm{blk}}
=
-\frac12\pi_{0,\sigma}'',
\label{eq:appC_pi20_blk}
\end{equation}
together with homogeneous boundary conditions.
Considering that
$\pi_{0,\sigma}''=O(1)$ and
$\mathcal A=\varepsilon\mathcal A^{(0)}$, we obtain
\begin{equation}
\pi_{20,\sigma}^{\rm blk}
=
O(\varepsilon^{-1}).
\label{eq:appC_pi20_blk_scaling}
\end{equation}
At the interval centre,
$\pi_{20,\sigma}^{\rm BL}=O(\varepsilon^{-1/2})$ and is
therefore smaller than the bulk contribution by a relative
factor $O(\sqrt{\varepsilon})$.

For the directional moment, we denote by
$m_{20,\sigma}^{\mathrm{blk}}$ the contribution generated
by the spatially extended sources
$-\pi_{20,\sigma}^{\mathrm{blk}}$ and
$-m_{0,\sigma}''/2$.  $m_{20,\sigma}^{\mathrm{blk}}$ obeys
\begin{equation}
\mathcal A m_{20,\sigma}^{\mathrm{blk}}
=
-\pi_{20,\sigma}^{\mathrm{blk}}
-\frac12m_{0,\sigma}'',
\label{eq:appC_m20_blk}
\end{equation}
again with homogeneous boundary conditions. Since both
$\pi_{20,\sigma}^{\rm blk}$ and
$m_{0,\sigma}''$ are $O(\varepsilon^{-1})$,
\begin{equation}
m_{20,\sigma}^{\rm blk}
=
O(\varepsilon^{-2}).
\label{eq:appC_m20_blk_scaling}
\end{equation}
The boundary layer generated contributions to
$m_{20,\sigma}(0)$ are instead
$O(\varepsilon^{-3/2})$ and are therefore smaller than
the bulk contribution by a relative factor
$O(\sqrt{\varepsilon})$.

Thus the leading rapid angular relaxation contribution to
the $O(U^2)$ conditional mean first passage time is determined by
$\pi_{20,\sigma}^{\rm blk}$ and
$m_{20,\sigma}^{\rm blk}$.
\subsection{Derivative representation of the bulk coefficients}
\label{appC:enhanced_diffusivity}

Having established that the boundary-layer contributions
are subleading, we now relate the bulk coefficients
$\pi_{20,\sigma}^{\mathrm{blk}}$ and
$m_{20,\sigma}^{\mathrm{blk}}$ to the corresponding
passive solutions. For this purpose, we reparametrise the
passive problem in terms of $\widetilde D$ and the drift
$v=\widetilde D F$. The bulk $O(U^2)$ equations can then
be related to derivatives of the passive solutions with
respect to $\widetilde D$ at fixed $v$.

\subsubsection{Splitting probability}

\noindent The passive splitting probability satisfies
Eq.~\eqref{eq:appC_pi0},
\begin{equation*}
\mathcal A\pi_{0,\sigma}=0.
\end{equation*}
Differentiating this equation with respect to
$\widetilde D$ at fixed $v$ gives
\begin{equation}
\mathcal A
\left(
\partial_{\widetilde D}\pi_{0,\sigma}
\right)_v
=
-\pi_{0,\sigma}''.
\label{eq:appC_pi0_dD_eq}
\end{equation}
Hence
\begin{equation}
\mathcal A
\left[
\frac12
\left(
\partial_{\widetilde D}\pi_{0,\sigma}
\right)_v
\right]
=
-\frac12\pi_{0,\sigma}''.
\label{eq:appC_pi0_dD_half}
\end{equation}
Comparison with
Eq.~\eqref{eq:appC_pi20_blk} shows that
$\frac12(\partial_{\widetilde D}\pi_{0,\sigma})_v$
and $\pi_{20,\sigma}^{\mathrm{blk}}$ satisfy the same
differential equation.
Their boundary conditions also coincide. The boundary
values of the passive splitting probability are fixed at
$0$ or $1$ and do not depend on $\widetilde D$.
Therefore,
\begin{equation}
\left.
\left(
\partial_{\widetilde D}\pi_{0,\sigma}
\right)_v
\right|_{X_0=\pm1/2}
=
0.
\label{eq:appC_pi0_dD_BC}
\end{equation}
Thus
$\frac12(\partial_{\widetilde D}\pi_{0,\sigma})_v$
satisfies both the same differential equation and the
same homogeneous boundary conditions as
$\pi_{20,\sigma}^{\mathrm{blk}}$. By uniqueness of the Dirichlet boundary-value problem,
\begin{equation}
\pi_{20,\sigma}^{\mathrm{blk}}
=
\frac12
\left(
\partial_{\widetilde D}\pi_{0,\sigma}
\right)_v.
\label{eq:appC_pi20_derivative}
\end{equation}

\subsubsection{First directional moment}

\noindent The passive first directional moment is governed by
Eq.~\eqref{eq:appC_m0},
\begin{equation*}
\mathcal A m_{0,\sigma}
=
-\pi_{0,\sigma}.
\end{equation*}
Differentiating with respect to $\widetilde D$ at fixed
$v$ results in
\begin{equation}
\mathcal A
\left(
\partial_{\widetilde D}m_{0,\sigma}
\right)_v
=
-
\left(
\partial_{\widetilde D}\pi_{0,\sigma}
\right)_v
-
m_{0,\sigma}''.
\label{eq:appC_m0_dD_eq}
\end{equation}
Using Eq.~\eqref{eq:appC_pi20_derivative}, we obtain
\begin{equation}
\mathcal A
\left[
\frac12
\left(
\partial_{\widetilde D}m_{0,\sigma}
\right)_v
\right]
=
-\pi_{20,\sigma}^{\mathrm{blk}}
-\frac12m_{0,\sigma}''.
\label{eq:appC_m0_dD_half}
\end{equation}
Comparison with Eq.~\eqref{eq:appC_m20_blk} shows that
$\frac12(\partial_{\widetilde D}m_{0,\sigma})_v$
and $m_{20,\sigma}^{\mathrm{blk}}$ satisfy the same
differential equation.
The passive directional moment vanishes at both
absorbing boundaries for every value of
$\widetilde D$, such that
\begin{equation}
\left.
\left(
\partial_{\widetilde D}m_{0,\sigma}
\right)_v
\right|_{X_0=\pm1/2}
=
0.
\label{eq:appC_m0_dD_BC}
\end{equation}
Thus
$\frac12(\partial_{\widetilde D}m_{0,\sigma})_v$
satisfies both the same differential equation and the
same homogeneous boundary conditions as
$m_{20,\sigma}^{\mathrm{blk}}$, we therefore obtain
\begin{equation}
m_{20,\sigma}^{\mathrm{blk}}
=
\frac12
\left(
\partial_{\widetilde D}m_{0,\sigma}
\right)_v.
\label{eq:appC_m20_derivative}
\end{equation}

Equations~\eqref{eq:appC_pi20_derivative} and
\eqref{eq:appC_m20_derivative} express the bulk
$O(U^2)$ coefficients as derivatives of the corresponding
passive solutions with respect to $\widetilde D$ at fixed
$v$.

\subsection{Leading correction to the direction-conditioned
mean first passage times}
\label{appC:conditional_mean}

After averaging over the initial orientation and setting
$X_0=0$, the splitting probability and first directional
moment have the expansions
\begin{align}
\Pi_\sigma
&=
\pi_{0,\sigma}(0)
+
U^2\pi_{20,\sigma}(0)
+
O(U^4),
\\
M_\sigma
&=
m_{0,\sigma}(0)
+
U^2m_{20,\sigma}(0)
+
O(U^4).
\end{align}
The corresponding direction-conditioned mean first
passage time,
$\langle\tau_\sigma\rangle=M_\sigma/\Pi_\sigma$, is
therefore
\begin{equation}
\langle\tau_\sigma\rangle
=
T_{0,\sigma}
+
U^2T_\sigma^{(2)}
+
O(U^4),
\label{eq:appC_mean_expansion}
\end{equation}
where
\begin{equation}
T_{0,\sigma}
=
\frac{m_{0,\sigma}(0)}
{\pi_{0,\sigma}(0)}
\label{eq:appC_T0_def}
\end{equation}
is the passive conditional mean first passage time, and
\begin{equation}
T_\sigma^{(2)}
=
\frac{
m_{20,\sigma}(0)
-
T_{0,\sigma}\pi_{20,\sigma}(0)}
{\pi_{0,\sigma}(0)}.
\label{eq:appC_T2_def}
\end{equation}

We first retain the bulk terms derived in the preceding
subsections. Their contribution to
Eq.~\eqref{eq:appC_T2_def} is
\begin{equation}
T_{\sigma,\mathrm{blk}}^{(2)}
=
\frac{
m_{20,\sigma}^{\mathrm{blk}}(0)
-
T_{0,\sigma}
\pi_{20,\sigma}^{\mathrm{blk}}(0)}
{\pi_{0,\sigma}(0)}.
\label{eq:appC_T2_blk_def}
\end{equation}
Using
Eqs.~\eqref{eq:appC_pi20_derivative} and
\eqref{eq:appC_m20_derivative}, the quotient rule applied
to Eq.~\eqref{eq:appC_T0_def} renders
\begin{equation}
T_{\sigma,\mathrm{blk}}^{(2)}
=
\frac12
\left(
\partial_{\widetilde D}T_{0,\sigma}
\right)_v .
\label{eq:appC_T2_derivative}
\end{equation}
For passive drift-diffusion from the interval centre, the
two direction-conditioned mean first passage times are
identical:
\begin{equation}
T_{0,\sigma}(\widetilde D,v)
=
\frac{1}{2v}
\tanh\!\left(
\frac{v}{4\widetilde D}
\right),
\qquad v\neq0,
\label{eq:appC_T0_passive}
\end{equation}
with the continuous unbiased limit
\begin{equation}
T_{0,\sigma}
=
\frac{1}{8\widetilde D},
\qquad v\to0.
\label{eq:appC_T0_passive_unbiased}
\end{equation}
Equation~\eqref{eq:appC_T2_derivative} therefore gives the
same bulk $O(U^2)$ coefficient for both exit directions,
\begin{equation}
T_{\sigma,\mathrm{blk}}^{(2)}
=
-\frac{1}{16\widetilde D^{\,2}}
\sech^2\!\left(
\frac{v}{4\widetilde D}
\right).
\label{eq:appC_T2_blk}
\end{equation}
Evaluating this expression at
$\widetilde D=\varepsilon$ and
$v=\varepsilon F$ we find
\begin{equation}
T_{\sigma,\mathrm{blk}}^{(2)}
=
-\frac{1}{16\varepsilon^2}
\sech^2\!\left(
\frac{F}{4}
\right).
\label{eq:appC_T2_blk_eps}
\end{equation}

The complete coefficient also contains the contributions
generated by the boundary layers. We therefore write
\begin{equation}
T_\sigma^{(2)}(\varepsilon,F)
=
-\frac{1}{16\varepsilon^2}
\sech^2\!\left(
\frac{F}{4}
\right)
+
R_\sigma(\varepsilon,F),
\label{eq:appC_T2_remainder}
\end{equation}
where $R_\sigma$ collects these subleading terms. The
estimates obtained above for
$\pi_{20,\sigma}(0)$ and $m_{20,\sigma}(0)$ show that
\begin{equation}
R_\sigma(\varepsilon,F)
=
O(\varepsilon^{-3/2}),
\qquad
\varepsilon\to0.
\label{eq:appC_R_scaling}
\end{equation}
This estimate applies to each exit direction separately
and need not be sharp for the difference $R_+-R_-$,
because leading contributions to the two remainders may
cancel. Since the bulk term in
Eq.~\eqref{eq:appC_T2_remainder} is common to both
directions,
\begin{equation*}
T_+^{(2)}-T_-^{(2)}
=
R_+-R_- .
\end{equation*}
Equation~\eqref{eq:appC_R_scaling} therefore implies
\begin{equation}
\varepsilon^2
\left[
T_+^{(2)}(\varepsilon,F)
-
T_-^{(2)}(\varepsilon,F)
\right]
\rightarrow0,
\qquad
\varepsilon\to0.
\label{eq:appC_T2_directional_difference}
\end{equation}
Combining the weak-persistence expansion with
Eq.~\eqref{eq:appC_T2_remainder} gives
\begin{equation}
\begin{aligned}
\langle\tau_\sigma\rangle
={}&
\frac{1}{2\varepsilon F}
\tanh\!\left(\frac{F}{4}\right)
\\
&+
U^2\left[
-\frac{1}{16\varepsilon^2}
\sech^2\!\left(\frac{F}{4}\right)
+
R_\sigma(\varepsilon,F)
\right]
+
O(U^4).
\end{aligned}
\label{eq:appC_mean_asymptotic}
\end{equation}

Equation~\eqref{eq:appC_T2_derivative} also gives a simple
interpretation of the common bulk correction in terms of
the passive problem. At fixed drift $v$,
\begin{equation}
T_{0,\sigma}(\widetilde D+\delta D,v)
=
T_{0,\sigma}(\widetilde D,v)
+
\delta D
\left(
\partial_{\widetilde D}T_{0,\sigma}
\right)_v
+
O(\delta D^2).
\label{eq:appC_T0_Dshift}
\end{equation}
Choosing
$\delta D=U^2/2$ and using $D_{\mathrm{eff}}$ defined in Eq.~\eqref{eq:Deff_wp} gives
\begin{equation}
T_{0,\sigma}(D_{\mathrm{eff}},v)
=
T_{0,\sigma}(\widetilde D,v)
+
U^2T_{\sigma,\mathrm{blk}}^{(2)}
+
O(U^4).
\label{eq:appC_T0_Deff}
\end{equation}
Thus the common bulk $O(U^2)$ correction is the change in
the passive conditional mean obtained by replacing
$\widetilde D$ with $D_{\mathrm{eff}}$ while keeping the
drift $v$ unchanged.

\begin{figure}[t]
    \centering
    \includegraphics[width=1.02\columnwidth]{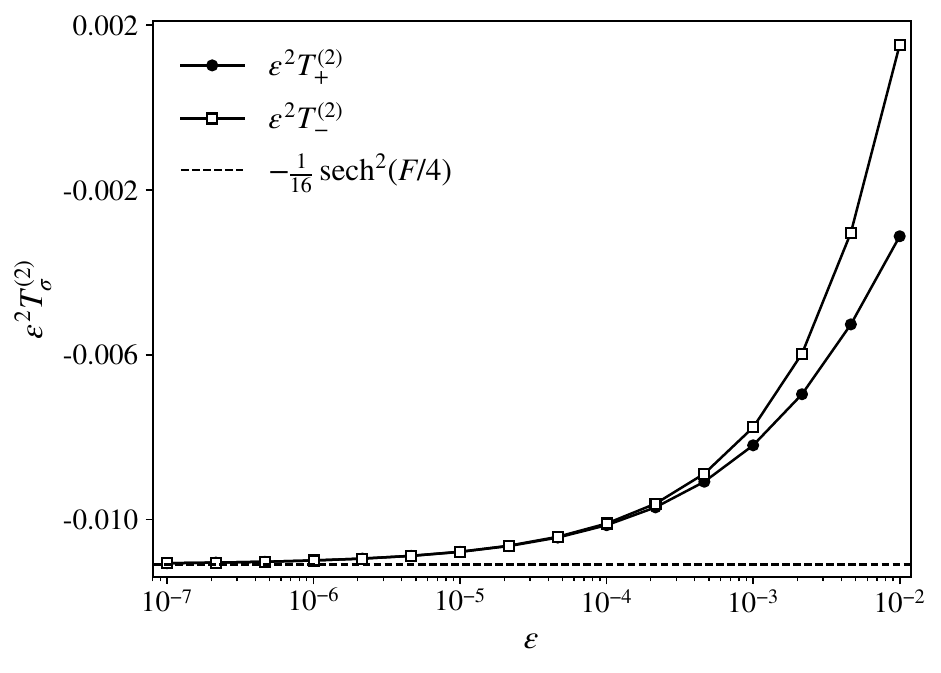}
    \caption{
  Numerical confirmation of the common leading asymptote
in the rapid angular relaxation limit at $F=6.04$. The rescaled coefficients
$\varepsilon^2T_{+}^{(2)}$ (black filled circles) and
$\varepsilon^2T_{-}^{(2)}$ (black open squares) were
obtained by solving the perturbative boundary-value
hierarchy in Eqs.~\eqref{eq:appC_pi_hierarchy} and
\eqref{eq:appC_m_hierarchy} using cubic ($p=3$)
Lagrange finite elements.
The dashed horizontal line denotes the common leading
asymptotic value
$-\frac{1}{16}\sech^2(F/4)$ from
Eq.~\eqref{eq:appC_T2_blk_eps}.
At the smallest value shown, $\varepsilon=10^{-7}$,
the numerical values are
$\varepsilon^2T_{+}^{(2)}=-0.01106061$ and
$\varepsilon^2T_{-}^{(2)}=-0.01106057$, compared with
the asymptotic value $-0.01109135$.
The finite $\varepsilon$ separation of the two curves is
contained in the difference between the subleading
remainders $R_+$ and $R_-$.
    }
    \label{fig:T2_validation}
\end{figure}
\subsection{Numerical confirmation of the common leading asymptote}
\label{sec:appendixC_numerics}

To test the asymptotic result against the full
perturbative hierarchy, we solved
Eqs.~\eqref{eq:appC_pi_hierarchy} and
\eqref{eq:appC_m_hierarchy} for $F=6.04$ and
$10^{-7}\leq\varepsilon\leq10^{-2}$.
The equations were discretised using cubic ($p=3$)
Lagrange finite elements on a nonuniform mesh, with
elements clustered near the absorbing boundaries
$X_0=\pm1/2$ to resolve the increasingly narrow boundary
layers as $\varepsilon$ decreases. The homogeneous
boundary conditions were imposed directly, and the
linear problems were solved in the order defined by the
perturbative hierarchy.

The results shown in Fig.~\ref{fig:T2_validation} were
obtained using $2400$ cubic elements. Comparison with a
calculation using $1200$ elements showed changes well
below the graphical resolution over the range of
$\varepsilon$ considered.

Figure~\ref{fig:T2_validation} shows that both
$\varepsilon^2T_{+}^{(2)}$ and
$\varepsilon^2T_{-}^{(2)}$ approach the common limiting
value
$-\frac{1}{16}\sech^2\!\left(\frac{F}{4}\right)$
as $\varepsilon\to0$, in agreement with
Eq.~\eqref{eq:appC_T2_blk_eps}. The separation between
the two curves decreases as $\varepsilon$ is reduced,
consistent with the subleading boundary layer
contributions identified above.


\begin{acknowledgments} 
A.R.\ gratefully acknowledges financial support from the Czech Science Foundation (Project No.\ 23-09074L). 
M.T. acknowledges financial support from the Portuguese Foundation for Science and Technology (FCT) under the contracts:  UID/00618/2025 (DOI: 10.54499/UID/00618/2025), UID/PRR2/00618/2025 (DOI: 10.54499/UID/PRR2/00618/2025), and UID/PRR/00618/2025 (DOI: 10.54499/UID/PRR/00618/2025). Some of the computational resources used in this work were provided by the e-INFRA CZ project (ID:90254), supported by the Ministry of Education, Youth and Sports of the Czech Republic. 
\end{acknowledgments}


%

\end{document}